%% file: ms.tex
\documentclass[12pt, preprint]{aastex}

\newcommand\um{\ifmmode{\mu{\rm m}}\else{$\mu$m}\fi}

\begin{document}
\title{Dust Emission from Unobscured Active Galactic Nuclei}
\author{G. D. Thompson, N. A. Levenson, S. A. Uddin, and M. M. Sirocky}
\affil{Department of Physics and Astronomy, University of Kentucky, Lexington, KY 40506; gdthom4@uky.edu}
\shortauthors{Thompson et al.}
\shorttitle{Dust Emission from Unobscured AGNs}
\begin{abstract}
We use mid-infrared spectroscopy of unobscured active galactic nuclei (AGNs) to reveal their native dusty environments.  We concentrate on Seyfert 1 galaxies, observing a sample of 31 with the Infrared Spectrograph aboard the \textit{Spitzer Space Telescope}, and compare them with 21 higher-luminosity quasar counterparts.  Silicate dust reprocessing dominates the mid-infrared spectra, and we generally measure the 10 and 18\um{} spectral features weakly in emission in these galaxies.  The strengths of the two silicate features together are sensitive to the dust distribution.  We present numerical radiative transfer calculations that distinguish between clumpy and smooth geometries, which are applicable to any central heating source, including stars as well as AGNs.  In the observations, we detect the obscuring ``torus'' of unified AGN schemes, modeling it as compact and clumpy.  We also determine that star formation increases with AGN luminosity, although the proportion of the galaxies' bolometric luminosity attributable to stars decreases with AGN luminosity.
\end{abstract}


\section{Introduction}
Accretion onto central supermassive ($10^{6}$-$10^{10}M_\odot$) black holes powers active galactic nuclei (AGNs).  AGNs exhibit a great variety of observational characteristics, notably the presence or absence of spectrally broad emission lines, which determine their classification as type 1 or type 2, respectively.  Unified AGN models \citep{antonucci} account for these differences in terms of viewing geometry, with an optically and geometrically thick dusty torus that blocks the broad line region and central engine from some (type 2) lines of sight.

The presence of the dusty torus can be detected not only in absorption but also in emission, with the bulk of the reprocessed AGN continuum emerging at infrared wavelengths.  The exact spectral energy distribution (SED) is a function of dust geometry, and initial radiative transfer calculations modeled a smooth, uniform torus, which is consistent with basic requirements of unification \citep{pierkrolik92,granato94,Row95}.  However, recent high-resolution observations indicate that the obscuring torus has a clumpy geometry \citep{jaffe04, tristram07, beckert08}. These data require a large range of dust temperatures to coexist at the same distance from the central heating source, whereas the temperature of a smooth torus declines with distance.  New models that place the dust in a clumpy toroidal distribution account for these observations while remaining consistent with unification schemes \citep{nenkova02,honig06,schartmann08}.

Silicate dust specifically dominates the mid-infrared (MIR) spectra of galaxies, and it can produce both the continuum and prominent spectral features at 10 and 18\um.  The stronger 10\um{} feature originates from a SiO stretching mode and the 18\um{} feature from a SiO bending mode \citep{knacke73}.  A view of a hot, optically thin surface produces the features in emission, and a view through a cold screen shows the features in absorption.  A smooth torus therefore exhibits strong emission from type 1 orientations (which view the hot inner throat of the torus directly), and deep absorption in type 2 AGNs (viewed through the torus).  However, observations of AGNs do not conform to these expectations.  Silicate emission is not universally observed in type 1 galaxies, and only recently has it been detected, primarily in high-luminosity AGNs \citep{hao05,hao07,siebenmorgen05,sturm05}.  Current observations of samples of the lower-luminosity Seyfert 1 galaxies even show silicate absorption on average \citep{hao07}, and Seyfert 2 galaxies typically show only weak absorption \citep{mason06,shi,hao07}.

Here we use MIR spectroscopy of unobscured AGNs to diagnose their native dusty environments.  Even more informative than the behavior of one silicate feature alone, the combination of both silicate features together reveals the geometry of the reprocessing dust around the AGNs, discriminating between smooth and clumpy distributions \citep{sirocky}.  We also use the spectra to evaluate the energetic contribution of star formation, which is evident in low-ionization emission lines and polycyclic aromatic hydrocarbon (PAH) emission, and which accounts for a significant fraction of the long-wavelength continuum flux.

\section{Sample and Data Reduction\label{sec:data}}
We select Seyfert 1 galaxies from the \citet{rush} 12\um{} survey, using 31 with archival low-resolution spectra from the \textit{Spitzer Space Telescope} Infrared Spectrograph (IRS) in this study \citep{Werner04,houck}.  The hot dust continuum of AGNs dominates the MIR emission in this flux-limited sample.  We restrict the Seyfert 1s to those galaxies with optical classification between 1 and 1.5.  As a comparison sample, we use 21 nearby ($z < 0.4$) type 1 quasars having archival IRS spectra.  We classify all sources with monochromatic luminosity ($\nu L_{\nu}$) at 14\um{} $L_{14}\geq 7 \times 10^{10} L_\odot$ as quasars.  We list the galaxies, their basic properties, and observational details in Table \ref{tab:sources}.

These IRS observations cover the MIR bandpass from 5.2 to 38\um{}.  We used standard \textit{Spitzer} Science Center pipeline version S15.3.0 data and extracted the spectra with the \textit{Spitzer} IRS Custom Extraction (SPICE) package.  The observations were performed in either staring or mapping mode.  We used two nodded positions of the same order to background subtract the staring mode observations, in which the nucleus is always centered in the slit.  Differencing exposures in first and second orders provided background subtraction of mapping mode data.  For these mapping mode observations, in addition to the central spectrum obtained with the slit centered on the nucleus, we included contributions from the two adjacent off-center spectra.  We compared the galaxies' FWHM to that of a calibration star to identify extended sources, finding eight: ESO 12-G21, MCG -5-13-17, NGC 7469, IC 4329A, NGC 1566, UGC 5101, NGC 3227, and MCG -6-30-15.  We used default SPICE extraction for point sources and extracted only the central 4 pixels (which corresponds to 7.2\arcsec{} and 20.4\arcsec{} in the short- and long-wavelength orders, respectively) from extended spectra.  We robustly averaged individual spectra from each order to remove bad data.  We scaled short wavelength spectra to match the flux of the long-wavelength order ``Long Low 1,'' which has the widest slit and is less sensitive to pointing errors.

\section{Results\label{sec:results}}
\subsection{Broadband Spectral Characteristics\label{sec:overall}}
The Seyfert 1 sample and the comparison quasars reveal the typical MIR characteristics of unobscured AGNs alone, which are evident in their average spectra (Figures~\ref{fig1} and \ref{fig2}).  The individual spectra are normalized at 14\um, which represents the dust continuum, and weighted according to their signal-to-noise in each average spectrum.  Both the Seyfert and quasar spectra show prominent 10\um{} silicate emission, which reveals the geometry of the dust distribution, high-ionization emission lines, which are predominantly the result of ionization by the AGN continuum, and polycyclic aromatic hydrocarbon (PAH) emission, which is associated with star formation.

A few sample members are luminous or ultraluminous infrared galaxies (LIRGs or ULIRGs), having $L_{IR} >10^{11} L_\odot$ or $>10^{12} L_\odot$, respectively.  These galaxies (plotted as red histograms) are preferentially mergers \citep{sanders96}, and they often exhibit extended MIR emission \citep{soi00, alo06}, which is not directly attributable to the AGN.  We show the average Seyfert 1 and quasar spectra, both including (red) and excluding (black) these (U)LIRGs.  The IR-luminous galaxies' spectra are different from the others', showing silicate absorption and relatively strong long-wavelength emission.  However, because only two of 31 Seyfert 1 galaxies and one of 21 quasars are also (U)LIRGs, the resulting average spectra are not significantly different in each case.  We base the subsequent analysis on the (U)LIRG-free average spectra, which better isolate the AGN contribution.  Table~\ref{table:avgspectra} contains the normalized average spectra of the Seyfert 1 galaxies and quasars.

Figure~\ref{fig2} further shows both the average Seyfert spectrum (green) and the average quasar spectrum (black) together.  Overall, these spectra are very similar in shape and emission features.  The 5 to 14\um{} flux density ratios, $F_{5}/F_{14}$, are slightly but not significantly different, with $F_{5}/F_{14} = 0.25\pm 0.12$ in the Seyfert 1s and $F_{5}/F _{14} = 0.45 \pm 0.16$ in the quasars.  We can describe the MIR spectra as a power law $F_{\nu} \propto\nu^{\alpha}$, where $\alpha < 0$ is characteristic of AGNs \citep{elvis, klaas, alonso}.  The power law indices range from -0.5 to -2.8 in the Seyfert 1 sample and -0.3 to -1.6 in the quasars, with $\alpha = -1.3$ and $-0.8$ in the average Seyfert 1 and quasar spectra, respectively.  These results agree with those of \citet{alonso}, who found $-2.8 < \alpha < -0.5$ over the wavelength range of 3.6--8\um{} in AGN-dominated galaxies.  

We also measured the equivalent widths (EWs) of strong emission lines in the average spectra.  We find the line EWs to be larger in the Seyfert galaxy spectra than in the quasar spectra, considering both strong high excitation lines, such as [\ion{O}{4}] and [\ion{Ne}{5}], which are attributable to AGNs \citep{lutz03}, and the low ionization lines [\ion{Ne}{2}] and [\ion{S}{3}], which originate in star formation.  Specifically, the [\ion{O}{4}] EW is 0.09\um{} in the average Seyfert spectrum and 0.03\um{} in the average quasar spectrum.  These results agree with previous MIR work by {\citet{honig08} and \citet{Keremedjiev08}.  \citet{Melendez} find $L_{[OIV]} \propto L_{2-10 keV}^{0.7}$, which similarly indicates smaller EW in the higher luminosity AGNs, if we consider the 2--10 keV X-ray luminosity as a proxy for intrinsic AGN luminosity.  Finally, while the quasar comparison sample was not selected robustly, we note the resulting average spectrum is extremely similar to that of \citet{netzer07}, showing a comparable spectral shape, broad silicate emission, and PAH emission.

\subsection{Type 1 AGN silicate emission\label{sec:emission}}

Dust produces both the MIR continuum and the silicate features.  We model the spectra assuming the same dust produces both, as opposed to invoking physically separate line- and continuum-producing regions \citep[c.f.][]{schweitzer08}.  In the IRS spectra, we measure the continuum over a short-wavelength region (typically 5--7\um), an intermediate point (around 14\um), and a long-wavelength region (typically 26.5--31.5\um), and we fit a spline to define the full continuum, utilizing the method of \citet{sirocky}.  The silicate features are evident in the dust absorption cross section, and the local minimum around 14\um{} produces the pseudo-continuum of the observed spectra at this wavelength.  The resulting continuum fits agree well with the radiative transfer model calculations of the emission from synthetic dust that lacks the silicate features \citep{sirocky}.  Figure~\ref{fig3} shows an example of the continuum fit to Mrk 766.  The procedure slightly varies depending on the spectral characteristics, with ``continuum-dominated spectra'' containing AGN emission lines and weak PAHs, ``PAH-dominated spectra'' exhibiting strong PAH emission, and ``absorption-dominated spectra'' showing strong ice and hydrocarbon absorption below 14\um{} \citep{spoon05,sirocky}.  Most of these spectra are continuum-dominated, which provide more reliable 10\um{} silicate measurements, while the silicate strength measurements of PAH-dominated and absorption-dominated spectra are more uncertain.

We measure the silicate feature strength \[S_{Sil}=\ln\frac{F_{obs}(\lambda)}{F_{cont}(\lambda)}\] at the wavelength of the strength extremum around 10 and 18\um, where $F_{obs}$ is the observed flux density and $F_{cont}$ is the fitted continuum flux density, as in \citet{lev}.  Table \ref{tab:observed} lists these silicate strengths ($S_{10}$ and $S_{18}$) and peak wavelengths ($\lambda_{10}$ and $\lambda_{18}$).  The average values of $\lambda_{10}$ and $\lambda_{18}$ are $10.0 \pm 0.1$\um{} and $18.1 \pm 0.2$\um{} in Seyfert 1 AGNs and $10.1 \pm 0.2$\um{} and $18.0 \pm 0.2$\um{} in quasars.  These values are consistent with the dust cross sections of \citet{ossenkopf} and characteristic of the interstellar medium, although radiative transfer effects can result in small  ($\lesssim 0.3$\um) wavelength shifts.  In contrast, we note that \citet{sturm05} attribute the large wavelength shifts they measure ($\lambda_{10}$ up to 11.5\um) to unusual grain size distributions.

The silicate strengths reveal features in emission ($S_{sil} > 0$) in 49 of 52 AGNs in the combined sample (Figure~\ref{fig4}), although the emission is generally weak.  The only type 1 AGNs that show 10\um{} silicate absorption are also LIRGs or ULIRGs.  The corresponding average Seyfert 1 spectrum shows the 10\um{} silicate feature in obvious emission ($S_{10}=0.11$), whereas the average Seyfert 1 spectrum of \citet{hao07} shows weak silicate absorption.  The homogeneous sample selection we employ and the exclusion of (U)LIRGs better isolates the AGN and its immediate environment in the MIR spectra.  In contrast, the heterogeneous Seyfert 1 sample of \citet{hao07} includes LIRGs and ULIRGs.  Most of the red 2MASS sources, for example, are indeed LIRGs and ULIRGS based on their IR luminosities.  The dusty star-forming regions of these (U)LIRGs contribute significantly to their IRS spectra and alter the appearance of the silicate features.  We conclude that the average Seyfert 1 spectrum of Figure~\ref{fig1}, which shows 10\um{} silicate emission, best characterizes the AGN and its immediate surroundings in the MIR.  The higher luminosity quasars similarly show 10\um{} emission, with $S_{10}=0.18$ on average.  Silicate emission in AGNs was first discovered in high luminosity galaxies \citep{hao05,siebenmorgen05,sturm05}, and although the quasar sample shows stronger 10\um{} silicate strength than the Seyfert sample, we find no significant correlation of silicate strength with AGN luminosity.  Finally, we find no trends with $\lambda_{10}$ or $\lambda_{18}$, indicating that the chemical composition of the dust does not vary significantly with AGN luminosity or silicate strength.

\subsection{AGN-Star Formation Connection\label{sec:starformation}}
Indicators of star formation, including [\ion{Ne}{2}] 12.8\um{} and several PAH bands \citep[and references therein]{Gen98}, are present in the majority of the spectra.  We measure the integrated luminosities of the [\ion{Ne}{2}] and 6.2\um{} PAH emission to quantify the star formation contribution, fitting a local continuum and a Gaussian for the line emission.  The 5\um{} monochromatic continuum luminosity scales with the  AGN luminosity, without the ambiguity of a star formation contribution to the continuum luminosity that is present at longer wavelengths.  These AGN and star formation luminosities are indeed positively correlated over both Seyfert 1 and quasar luminosities (Figure~\ref{fig5}).  We certainly measure [\ion{Ne}{2}] in all but two Seyfert galaxies, and the resulting robust [\ion{Ne}{2}]-5\um{} correlation (dashed line) is consistent with the less complete quasar measurements.  The [\ion{Ne}{2}] non-detections are a consequence of poor signal-to-noise.  Combining the spectra without detections, we successfully measure the line.  We plot these sources (green) at this average [\ion{Ne}{2}]-5\um{} ratio, considering the Seyfert 1 galaxies and quasars separately.  The solid line shows the subsequent correlation over all galaxies, including those in which [\ion{Ne}{2}] is not detected directly, which agrees with the Seyfert 1 result using individual detections alone.

The PAH measurements indicate similar trends of star formation increasing with AGN luminosity, but these results are less robust.  We directly detect the PAH emission in fewer individual spectra (24 of 31 Seyfert 1s and 8 of 21 quasars).  While we recover the PAH emission in the average ``PAH-less'' Seyfert 1 spectrum, we do not certainly detect PAH emission in the average ``PAH-less'' quasar spectrum.  Overall, this work agrees with that of \citet{schweitzer06} who find a similar correlation between AGN and starburst luminosity, measuring 6\um{} continuum luminosity and PAHs in quasars.   

While luminosity due to star formation and accretion are correlated, the relative contribution of star formation decreases with increasing AGN luminosity.  The slopes of the linear fits are $0.50 \pm 0.16$ and $0.66 \pm 0.07$ for the well-measured Seyferts and all sources, respectively.  Using equation 1 of \citet{hoketo}, we calculate the luminosity contribution of the starburst component from the [\ion{Ne}{2}] integrated luminosity, treating the infrared luminosity as an approximation of bolometric luminosity.  Similarly, using the median SED and bolometric corrections of \citet{elvis}, we obtain the bolometric scale factor for 5\um{} continuum, ${L_{bol}}=10.1 {L_{5}}$.  For weaker AGN contributors (e.g., $L_{5}=10^{8.5} L_{\odot}$), the star formation luminosity is as much as 80\% of the AGN contribution, whereas the star formation luminosity of strong quasars ($L_{5}=10^{12} L_{\odot}$) is around $ 5\%$ of their AGN luminosity.  The contribution of star formation to the continuum flux increases with wavelength.  Comparing the AGN-dominated 14\um{} flux density with that at 30\um, at which dust-reprocessed stellar light becomes significant, we find the average $F_{14}/F_{30} = 0.44\pm 0.18$ for the Seyfert 1s and $F_{14}/F_{30} = 0.68 \pm 0.27$ for the quasars.  Similar to the conclusions based on [\ion{Ne}{2}] and PAH emission, these results also suggest that the relative luminosity of star formation is greater in the Seyfert 1 galaxies than in the quasars.

\section{Dust Geometry from Silicate Features\label{sec:silicates}}

The strengths of the 10 and 18\um{} silicate features together are sensitive to the distribution of dust surrounding any heating source, including stars as well as AGNs.  Smooth and clumpy distributions occupy distinct regions of the ``feature-feature diagram'' \citep{sirocky}, which shows $S_{18}$ \textit{vs.} $S_{10}$ (Figure~\ref{fig6}).  We compare these radiative transfer calculations with the observations to discern the dusty environment of the AGNs we observed.  While the two strength measurements in an individual galaxy cannot constrain all the free parameters of any of the models, the type 1 AGNs are located in an area of the diagram that only clumpy models occupy.

\citet{schweitzer08} alternatively model MIR spectra of AGNs with multiple independent emission components.  Combinations of blackbodies represent the continuum, due to hot dust close to the nucleus. The more distant, cooler, optically-thin narrow line region, located at 100--200 dust sublimation radii produces the silicate features.  A disadvantage of this approach is that it allows silicate only in emission.  It never produces silicate absorption, which is typical of type 2 AGNs \citep{hao07}.  Indeed, assuming AGN unification, these models would instead predict silicate emission from type 2 AGNs in general.  Moreover, the hot inner edge of the disk \citet{schweitzer08} describe would also produce strong silicate emission that would be observed directly in unobscured AGNs but is absent from their model.  The key difference between the description of \citet{schweitzer08} and ours is the location of the silicate-emitting region.  Unfortunately, no current observations provide the spatial resolution to discriminate between them directly.  Thus, we pursue here models in which a common dust distribution simultaneously accounts for the MIR continuum and spectral features.  This technique has the advantage that it may be consistently and directly applicable to obscured AGNs, in agreement with unified AGN schemes, although we acknowledge that additional separate emission regions may be present in some galactic centers.

Independent of geometry, all optically thin configurations exhibit silicate emission and are located at the same point in the diagram, which corresponds to the feature strength in the optical cross section.  Tracks of increasing optical depth move toward weaker emission.  Only smooth dust distributions can exhibit large negative feature strengths, producing the temperature gradient that is essential for deep absorption.  All smooth spherical distributions show similar slopes in the diagram, independent of the density distribution and total dust extent, which the dust's optical properties determine \citep{sirocky}.  We plot several characteristic examples in Figure~\ref{fig6}.  Spherical distributions of clumps occupy a distinct region of the feature-feature diagram, never showing deep absorption, even for comparable total optical depth.  The reason for this behavior is that both dark (absorbed) and bright (illuminated) cloud faces are observed in the clumpy distribution.  The silicate emission from the bright sides fills in the absorption trough, reducing its depth \citep{nenkova02}.  

We consider whether changes to the dust composition could provide smooth distributions that describe the observations on the feature-feature diagram.  The dust's optical properties determine both the location of the optically thin point and the slopes of the smooth model tracks on the diagram \citep{sirocky}.  Compared to the dust we employ here \citep{ossenkopf}, the ``astronomical silicate'' of \citet{drainea,draineb}, for example, shows a reduced 18\um{} feature relative to the 10\um{} feature, so the silicate features of these smooth models (Figure 9 of Sirocky et al. 2008) are more similar to those of the observations.  However, the specific models that lie close to the data are optically thick, with $\tau_V \gg 10$, which is inconsistent with the small optical depths measured in type 1 AGNs.  This resulting optical thickness is a general problem for smooth models of any dust that exhibits the MIR silicate features. Only a contrived dust having extremely weak intrinsic silicate features could remain optically thin in the observed region of the diagram, yet even such a forced solution could not then produce the observed range of strength ratios.  We conclude that the data lie below the smooth model tracks for reasonable dust properties, a region that only clumpy models occupy.

While the smooth spherical models offer a valuable contrast, they are inappropriate in these cases, failing to allow direct views of the central engine and the resulting spectrally broad emission lines.  Nevertheless, the smooth spherical geometries do usefully indicate some of the realm of the smooth torus of classical AGN unification schemes, which provide unobscured views along select lines of sight.  Specifically, the obscured type 2 view through the torus is analogous to the spherical geometry, admitting no view of directly-illuminated hot dust.  Strictly unobscured (type 1) lines of sight to the central engine view the silicate emission from the optically thin illuminated surface of the torus directly.  The net result is silicate emission, located near the optically thin point of the feature-feature diagram, independent of the total dust optical depth in the torus, with only a weak absorption contribution from the cooler torus interior within the observing beam \citep{granato94, Row95, van03}.  In type 1 views, the silicate strength is sensitive to the shape of the inner edge of the dusty torus.  The models of \citet{pierkrolik92}, for example, do not typically exhibit strong silicate emission.  However, several independent lines of evidence argue against smooth dust distributions generally.  First, the MIR emission of AGNs is observed to be effectively isotropic \citep{lutz04, buchanan06, horst06}, whereas all smooth torus models predict type 1 AGNs to be significantly brighter than type 2 AGNs in the MIR, for a fixed intrinsic luminosity.  Second, all optically thick smooth descriptions produce very deep absorption in obscured nuclei, which is not generally 
observed \citep{hao07}.  

Thus, we pursue clumpy dust distributions, which \citet{krolikbegelman88} originally proposed.  The clumpy geometry allows a large range of dust temperatures to coexist at the same distance, as opposed to the monotonic temperature decline with distance that is characteristic of smooth models.  Interferometric observations of NGC 1068 with the VLTI, for example, resolve the 10\um{} emission and indicate cool dust located close to the nucleus \citep{jaffe04}.  Similarly, VLTI observations of the Circinus galaxy and NGC 3783 provide further evidence of a clumpy dust structure \citep{tristram07,beckert08}.  

Initially, we minimize the number of model free parameters and consider a spherical distribution of clumps.  Although this geometry does not generally provide any clear lines of sight to the central engine, the simplified spherical models are powerful, and they capture the essence of the MIR spectra.  Fundamentally, the total population of clouds produces the observed MIR emission, and the SEDs are insensitive to viewing orientation effects in any case, even when the dust distribution is not spherically symmetric.  (We explicitly demonstrate this result below, presenting calculations for toroidal distributions of clouds.)  We follow the formalism of \citet{sirocky}, which is based on the radiative transfer code of \citet{Nenkova08a}.  In the computations, the individual clouds are distributed according to Poisson statistics, with an average number of clouds along a radial ray, $N_{0}$.  The clumps are radially distributed according to a power law, $\propto r^{-q}$, from the dust sublimation radius, $R_d$ to an outer radius $R_o$, which we parameterize with $Y={R_{o}}/{R_{d}}$.  The bolometric luminosity of the central source sets $R_d$, with $R_d \simeq 0.4 (L_{bol}/10^{45} \mathrm{erg\ s^{-1}})^{1/2} $ for the AGN heating spectrum and dust sublimation temperature of 1500 K.  The optical depth of each cloud in the $V$ band is $\tau_V$, so the total average optical depth through all clouds is $N_0\tau_V$.  The dust includes both silicates and graphite, and we use the \citet{ossenkopf} cool silicate optical properties.  The model results we present are applicable to any heating source, not restricted to AGNs, because the dust erases all signatures of the incident spectrum from the emergent MIR emission.

We initially leave the model parameters unconstrained in order to show the effects different parameters have on the model curves in the feature-feature diagram.  The plotted simulations therefore do not all represent best- (or even ``good-'') fitting models.  In exploring these parameter variations, we will identify the parameter values that produce models that generally agree with the observations on the feature-feature diagram.

Figure~\ref{fig6} shows the silicate strengths of the clumpy sphere models for a range of $N_{0}$, with $Y=30$ and $q=1$. The optical depth per cloud increases along each track of fixed $N_0$, from $\tau_V = 10$ (at the upper right) to 80.  Overall, increasing $N_0$ results in diminished silicate emission, with silicate absorption emerging for $N_0>3$.  Having more clouds increases the chance that bright faces are obscured, and views of dark, absorbed faces occupy more lines of sight.  As the optical depth per cloud increases along the constant $N_0$ track, silicate strength initially decreases.  Around $\tau_V = 60$, which corresponds to $\tau_{10}\approx 3$, optical depth effects within individual clouds become important, and the 10\um{} silicate strength increases as the optical depth per cloud increases further.  The 18\um{} strength generally continues to decrease, with $\tau_{18}< 2$ per cloud when $\tau_V = 80$.  These clumpy models describe the type 1 AGN data well, typically with a small number of clouds ($N_{0}\sim2$).

The MIR-emitting region is compact, and clouds located far from the AGN do not significantly affect the silicate feature strengths.  We demonstrate this result first considering the radial density profile $q=2$ models over a range of outer size, $Y$ (Figure~\ref{fig7}).  Having steep radial density profiles, these distributions are inherently compact and therefore are not sensitive to the outer extent.  For example, $80\%$ of clouds are located within $3.6 R_{d}$ and $4.8 R_{d}$ for $Y=10$ and $Y=100$, respectively.  However, the shallower radial distributions are sensitive to the total size because the number of {\em nearby} clouds is a function of both $Y$ and $N_0$.  In these cases, replicating the MIR behavior of the small-$N_0$ compact distributions requires increasing $N_0$ as the total size increases. 

Small numbers of clouds along radial rays in spherical clumpy models best match the silicate strengths of type 1 data, with $N_0\sim2$.  We find little variation in silicate strength as a function of other parameters when $N_0$ is small, so the conclusion that the immediate surroundings of these AGNs contain few clouds along radial rays is robust.  However, for larger values of $N_0$, the silicate strengths depend sensitively on the combination of all parameters, including $q$ and $Y$, as well.

While the clumpy sphere captures the essence of the MIR emission from these AGNs, general support for unified AGN models, especially the requirement of unobscured lines of sight to the central engine, favors a toroidal distribution.  We use the clumpy toroidal models of \citet{nenkova02,Nenkova08a,nenkova08}, which allow variation of the viewing inclination angle, $i$, and torus scale height, $\sigma$, in addition to the parameters of the spherical model.  Figure~\ref{fig8} illustrates the clumpy torus.  The quantity $N_0$ now represents the average number of clouds through an equatorial ray of the torus, and we consider Gaussian distributions, where the average number of clouds $N_{los}(\beta) = N_0 \exp(-\beta^2/\sigma^2)$ along angle $\beta$ measured from the equator.  (The inclination angle is measured from the symmetry axis of the torus, so $\beta=90^{\circ}-i$.)  Unobscured views are more likely for small values of $i$, with the photon escape probability $P_{esc}=\exp(-N_{los}(\beta))$ describing the likelihood of an unobscured view of the central engine \citep{nenkova08}.

The clumpy nature of the dust is fundamental, and as a result, the toroidal distributions occupy a region of the feature-feature diagram similar to that of the spherical arrangements.  The toroidal calculations uphold the general conclusion that the MIR-emitting region is small.  For example, these inherently-compact $q=2$ results are insensitive to the total outer extent of the torus, similar to the spherical calculations above.  The flat ($q=0$) radial density profile, however, is a strong function of outer radius (Figure~\ref{fig9}).  In this case, the most compact tori (having $Y=10$) generally describe the data well, with $N_0\sim 4$.  The clouds are spread over a large volume in the extended ($Y=100$ or $Y=30$) tori when $q=0$.  With this radial distribution few clouds are located close to the nucleus, even when the total number of clouds is large.  For example, using model parameters $Y=100$, $q=0$, and $N_0=10$, an average of only two clouds are located within $20 R_d$ along equatorial rays.  As a result, this particular combination of parameters describes many of the observations.  However, a significant fraction of observations lie above these models (having greater $S_{18}$), even for large values of $N_0$.  No change in other parameters can shift the $q=0$, $Y=100$ model tracks up to account for the stronger $S_{18}$ measurements.  Thus, this combination of parameter values is not generally characteristic of the observed sample, although it may describe particular galaxies.

Computations of varying $q$ for fixed $Y$ again show that the nearby clouds determine the silicate feature strengths.  In Figure~\ref{fig10}, we plot the strengths for models in which $q$ ranges from 0 to 2, with $Y=30, i=30^{\circ}, \sigma=45^{\circ}$, and $N_0=1$, 2, 4, and 10.  Models having steeper density distributions require fewer clouds along radial rays to match the silicate strength of comparable models having shallower radial distributions.  Furthermore, as $q$ increases, a larger fraction of the clouds are located closer to the AGN.  For example, with $Y=30$, in a $q=0$ distribution, $80\%$ of all clouds are located within $24 R_d$ whereas the same fraction of clouds are confined to $4.4 R_d$ in the $q=2$ density profile.  The shallowest density profile ($q=0$) does not follow the general trend of decreasing strength with increasing $N_0$ as rapidly as the steeper distributions do.  Considering the observations, few clouds ($\lesssim 6$) along radial rays within a compact torus (size $\sim 15R_d$) accounts for the MIR silicate features of unobscured AGNs.

Increasing the torus scale height reduces the silicate strength for a given $N_0$, as Figure~\ref{fig11} shows.  The total cloud distribution determines the behavior of the MIR emission, and an increase in $\sigma$ results in more clouds overall for a given $N_{0}$.  Formally, $N_0$ sets only the average number of clouds along equatorial rays, and $\sigma$ determines how rapidly the radial number declines with altitude.  Again, with more clouds present, the bright cloud faces are more likely to be obscured, which reduces the net silicate emission and produces absorption in some cases.  Here, the increased total number of clouds ($N_{tot}$) is a consequence of increasing $\sigma$ rather than increasing $N_0$ alone.  However, even for $N_{0}=10$, the $\sigma=15^{\circ}$ model shows emission, because $N_{tot}$ is small in this narrow torus.  Few high-altitude clouds are present to block views of the directly-heated cloud surfaces that exhibit silicate emission.  

The dominant direct view of a hot optically thin surface produces strong silicate emission in general, which is typical of smooth torus models of type 1 AGNs.  The small-$\sigma$ case approaches a two-dimensional dust distribution and explicitly demonstrates the failure of this simplification (in a clumpy or smooth arrangement), given the observed weak silicate features.  Instead, the mixture of contributions from hot and cold cloud faces is the essence of the MIR emission.  Thus, despite the defect of the spherical models in not providing unobscured lines of sight to the AGN in general, they better approximate the MIR results than a two-dimensional geometry does, and the development to the toroidal configuration ultimately corrects this problem.

We consider several inclination angles for fixed $N_0$ and $\sigma$ and find no significant differences in the models' silicate strengths for small $i$ typical of type 1 views ($i\leq40^{\circ}$).  Thus, the silicate features do not usefully diagnose the viewing angle, and we adopt $i=30^{\circ}$ in the comparisons below.  According to unification schemes, the only difference between type 1 and type 2 AGNs is the viewing angle, whereby direct views of the central engines of type 2 AGNs are obscured, although the dusty AGN surroundings are inherently the same in both cases.  A smooth torus strictly separates the different types at a particular viewing angle, distinguishing lines of sight through the dusty torus material from unobscured views near the symmetry axis.  The clumpy formalism presents no strict dividing line, however.  Instead, type 1 or type 2 views may arise from any angle, but lines of sight near the equatorial plane are more likely to be obscured, and those near the symmetry axis are more likely to remain clear.  The escape probability, $P_{esc}$, describes the likelihood of a type 1 view, and it is a function of $i$, $\sigma$, and $N_0$.

We compare clumpy torus properties that describe the type 1 observations well to predict the silicate characteristics of their type 2 counterparts, viewed at higher inclination, which are obscured.  We find that these type 2 AGNs show less silicate emission than type 1 views of the same dust distribution.  Specifically, we calculate the silicate strengths for combinations of model parameters that describe the type 1 measurements well: $N_0 = 2$--4, $\tau_V=30$--60, $\sigma=30$--$60^{\circ}$,  with $Y=10$--30 for $q=1$, and $Y=10$--100 for $q=2$.  The corresponding type 2 views ($i=70^{\circ}$) yield $-0.4<S_{10}<0.15$ and $-0.15<S_{18}<0.18$, a wider range of silicate strength than the $i=30^\circ$ views exhibit.  Overall, the silicate is generally weakly absorbed in the obscured AGNs, which agrees with previous observations \citep[Levenson et al. 2009, in preparation]{hao07}.  Another way to test unified AGN schemes is to model the characteristic clumpy distribution of observed type 2 AGNs and compare this result with the type 1 distributions.  The peak of the 10\um{} silicate strength distribution of the \citet{hao07} Seyfert 2s ranges over $-0.4\leq S_{10}\leq -0.1$, which corresponds to $N_0\simeq 3$--4 in the $i=70^{\circ}$ clumpy torus models using the same parameter combinations as above.  Changing inclination to $i=30^{\circ}$ and holding all other parameters constant, the silicate strengths of the $N_0=3$ or 4 models match those of the type 1 observations.  These results are again consistent with standard AGN unification, whereby the central engine and its immediate environment are the same in all AGNs, and variations in viewing geometry alone produce observable differences.  One further consequence of the clumpy geometry is that it {\em can} produce silicate emission even when the AGN is obscured, which has been observed \citep{sturm06, teplitz06, hao07}, unlike a smooth distribution, in which type 2 AGNs always exhibit silicate absorption.  Even when the AGN and broad line region are hidden, direct view of some hot cloud faces can result in measurable silicate emission.  

Despite the model degeneracies, we identify several ranges of ``standard'' parameters that describe the data well and do not impose severe restrictions on other parameters.  In particular, we favor $Y=30$, $q=1$, $\sigma=45^{\circ}$, $i=30^{\circ}$, $\tau_V=30$--60, and $N_{0} \leq 6$.  Because the clouds close to the AGN determine the MIR emission, confining the total extent of the torus ($Y=10$ or 30) generally produces models that describe the data well.  Recent observations also show a compact MIR torus, with sizes of $3$--$5 R_d$ and $10 R_d$, in \citet{mason06} and \citet{tristram07}, respectively.  The extended ($Y=100$) torus is successful only when the clouds are concentrated (with radial density profile $q=2$).  Similarly, in the constant density distribution ($q=0$), only $Y=10$ yields acceptable results.  

The MIR measurements alone do not strongly constrain the torus scale height.  Instead, relative numbers of type 1 and 2 Seyfert galaxies indicate $\sigma \simeq 30$ to $45^{\circ}$ \citep{Schmitt01,hao05}, and consideration of the SED and $10\um${} silicate feature favors $\sigma = 30^{\circ}$ \citep{nenkova08}.  However, the $\sigma = 30^{\circ}$ models underpredict $S_{18}$ at lower values of $S_{10}$, even with large numbers of clouds ($N_0 > 10$).  Thus, we adopt $\sigma = 45^{\circ}$ as the standard value and consider the range $30^{\circ}<\sigma<60^{\circ}$ to be applicable to various individual galaxies.  A disk-like geometry suggested by \citet{schweitzer08} with a clumpy distribution could be identified as a small-$\sigma$ torus, but low $\sigma$ values disagree with the observations.  Because varying inclination angle does not significantly change the strength measurements provided that $i<40^{\circ}$, we discount directly face-on views and identify $20^{\circ}<i<40^{\circ}$ to be typical of these type 1 AGNs.  Furthermore, models of $\tau_V=30$--60 produce silicate strengths that are similar to those of the observations.

The physical characteristics of clumpy dust distributions are fundamentally different from those of smooth distributions.  Both spherical and toroidal clumpy models cover the same regions of the feature-feature diagram, which remain inaccessible to all smooth descriptions.  The behavior of the silicate features in the simplified clumpy spherical models yields results that are directly applicable to the more realistic clumpy torus models.  To account for the MIR observations, the small-scale dust distribution is relevant, and both geometries show how the total torus extent and radial distribution together govern the effective compactness.  In addition, small numbers of clouds within the small sphere or torus agree with the observations of type 1 AGNs.  However, the $N_0$ parameter is not constant across all models that describe the data: it increases from the sphere to the torus generally, and it increases with decreasing torus scale height.  We therefore conclude that the total number of clouds available to reprocess the intrinsic AGN flux, $N_{tot}$ (not $N_0$), and their distribution ultimately determine the behavior of the emergent MIR emission.  

Figure~\ref{fig12} demonstrates that for the same inputs of $Y$ and $q$, fitting the data requires larger values of $N_{0}$ in the torus than in the sphere.  In general, translating from any torus to sphere model requires $N_0(torus)>N_0(sphere)$.  The parameter $N_0$ describes only the number of clouds along the equatorial ray of the torus, and the number of clouds diminishes with altitude.  For a given $N_0$, the spherical model contains more clouds in total than the toroidal model does.  Thus, in order to achieve the same $N_{tot}$ in both geometries, $N_{0}$ must be greater in the torus.  The spherical models yield robust conclusions about the nature of the dust distribution around AGNs, confirming that few clouds within a small radius account for the observed MIR emission.  However, because the spherical geometries generally fail to provide an unobscured view of the central engine, which these type 1 AGNs demand, we conclude that a clumpy, dusty torus characterizes the immediate surroundings of AGNs.

\section{Conclusions}
Dust reprocesses the AGN continuum to emerge at MIR wavelengths, and we model both the resulting continuum and spectral features at 10 and 18\um{} due to a common dusty region.  Isolating unobscured AGNs, we find these features in emission, both in Seyfert 1 galaxies and in quasars.  The emission is weak, however, with average emission strength $S_{10}=0.11$ and 0.18 in the Seyferts and quasars, respectively.  In contrast, an optically thin medium, such as the directly-viewed hot interior of a smooth torus, would yield stronger emission ($S_{10}=1.2$).  We conclude that the observable but weak emission is a consequence of clumpy AGN surroundings.  We measure the peak wavelength $\lambda_{10}=10.0 \pm 0.1$ and $10.1 \pm 0.2$\um{} in the Seyfert 1s and quasars, respectively.  These values are consistent with radiative transfer computations using the optical properties of silicates that \citet{ossenkopf} model.  In agreement with earlier work, we find that star formation increases with AGN luminosity, using [\ion{Ne}{2}] and 6.2\um{} PAH to quantify the star formation contribution and the 5 and 14\um{} continua as proxies for the AGN luminosity.  The fractional contribution of star formation to the total bolometric luminosity of these galactic centers decreases with AGN luminosity.  Furthermore, considering the flux ratio $F_{5}/F_{14}$, we can describe the type 1 AGN spectra as a power law, with power law indices ranging from -0.5 to -2.8 in the Seyfert 1s and -0.3 to -1.6 in the 
quasars.

The strengths of the 10 and 18\um{} silicate features together are sensitive to the distribution of dust surrounding any heating source, including stars as well as AGNs.  We interpret the silicate strength measurements of these isolated AGNs as a consequence of a native AGN environment that is clumpy.  The dust that determines the MIR behavior is confined to small scales.  In radiative transfer calculations, either limiting the total radial extent of the dust or concentrating the cloud distribution effectively produces the compact distributions that describe the observations.

A toroidal distribution of clouds is consistent with unified AGN schemes, offering unobscured lines of sight to the central engine from some viewing angles.  The total cloud distribution, not only the clouds located along the line of sight, determines the MIR emission.  As a consequence, the MIR output is effectively isotropic, as observations of all types of AGNs show \citep[e.g.,][]{horst06}.  We thus account for the model results in general: properties such as torus scale height and the number of clouds along radial rays that determine the total cloud distribution govern the MIR behavior.  Although spherical distributions do not generally provide the unobscured views these type 1 AGNs require, they do usefully capture the essence of the total cloud distribution while minimizing the number of free parameters.  While two silicate strength measurements cannot constrain all the free parameters of any of the models, we find that the type 1 AGNs are located in an area of the feature-feature diagram that only clumpy models occupy.  Few clouds are located along radial rays within the compact torus, which is consistent with column density variability observed in some AGNs \citep{Elv04,Ris07}.  The MIR observations are not sensitive to the more distant cloud population, and we conclude the arrangement of dust immediately surrounding the AGN central engine is a clumpy torus that contains few clouds ($\lesssim 6$) along radial rays within a small radius ($\sim 15R_d$).

\acknowledgements
We thank the anonymous referee for a useful review that improved this paper.  This work is based on observations made with the Spitzer Space Telescope, which is operated by the Jet Propulsion Laboratory, California Institute of Technology under a contract with NASA.  We acknowledge work supported by the NSF under Grant 0237291.

\textit{Facility:} \facility{Spitzer (IRS)}

\input tab1

\input tab2
\input tab3

\clearpage
\begin{figure}[htb!]
\centerline{\includegraphics[width=6in]{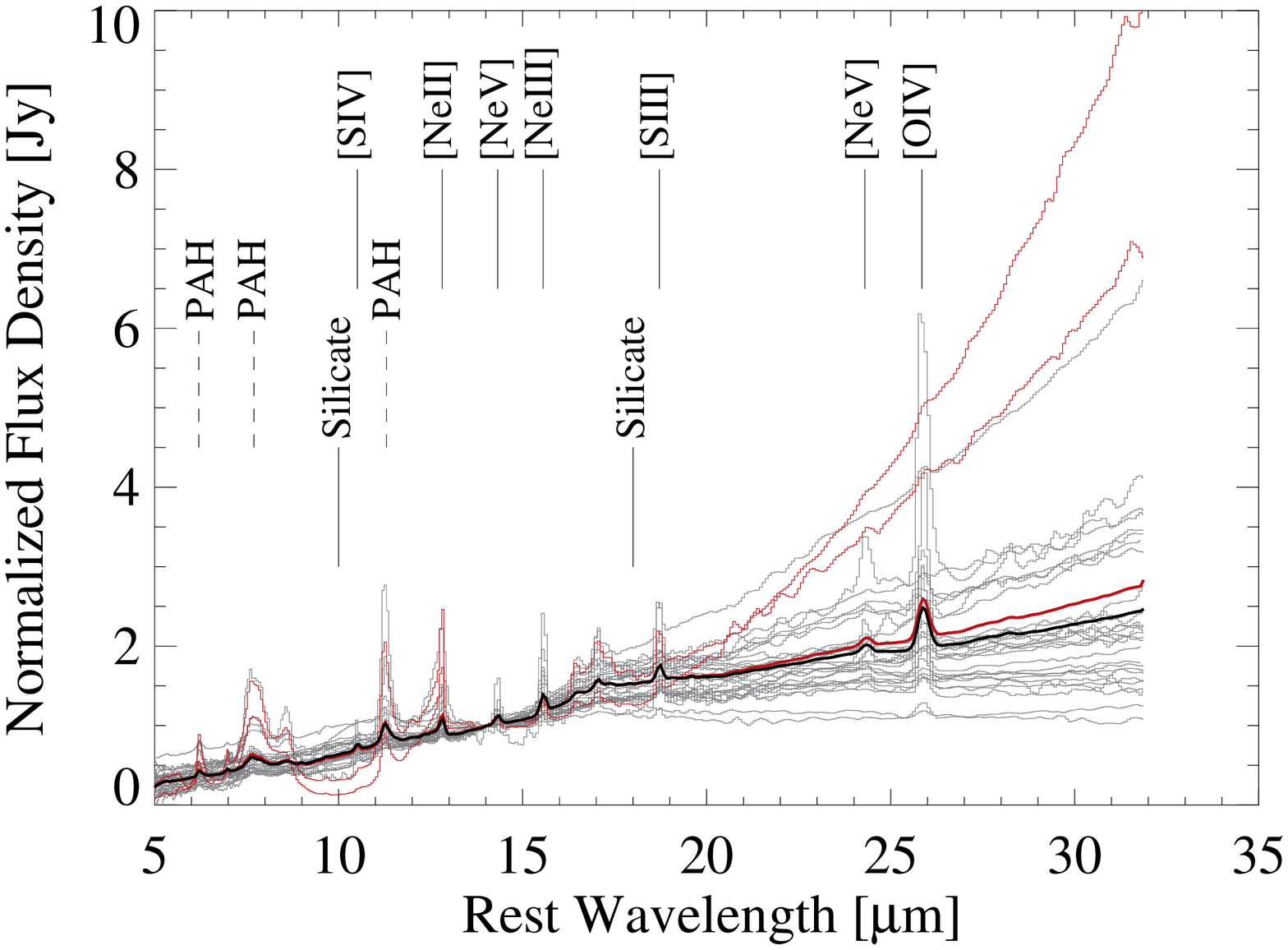}}
\caption{\label{fig1} Average Seyfert 1 spectra (smooth heavy lines) show 10 and 18\um{} silicate emission.  The bold black line best isolates the AGNs alone and is the average of 29 individual sources (grey histograms, normalized at 14\um) excluding LIRGs UGC 5101 and Mrk 1034 (red histograms).  The bold red line is the average spectrum of all sources.  In addition to the silicates, emission features that are characteristic of both AGNs and star formation are labeled.}
\end{figure}

\clearpage
\begin{figure}[htb!]
\centerline{\includegraphics[width=6in]{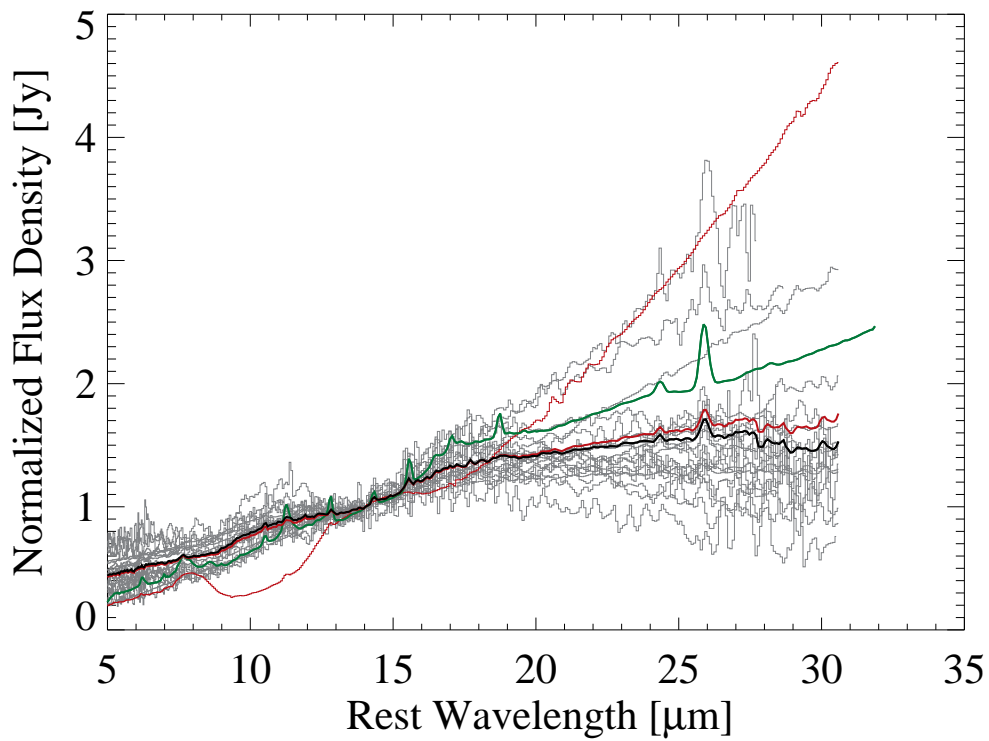}}
\caption{\label{fig2} Average quasar spectra (smooth heavy lines) show 10 and 18\um{} silicate emission.  The spectrum plotted in black best isolates the MIR emission of these luminous AGNs and excludes the ULIRG Mrk 231 (red histogram), which contributes to the average plotted in red.  As a comparison, the average Seyfert 1 spectrum is overplotted in green.  The equivalent widths of both AGN-originating emission lines, such as [\ion{O}{4}], and star formation indicators, such as PAHs and [\ion{Ne}{2}], are larger in the average Seyfert 1 spectrum.}
\end{figure}

\clearpage
\begin{figure}[htb!]
\centerline{\includegraphics[width=6in]{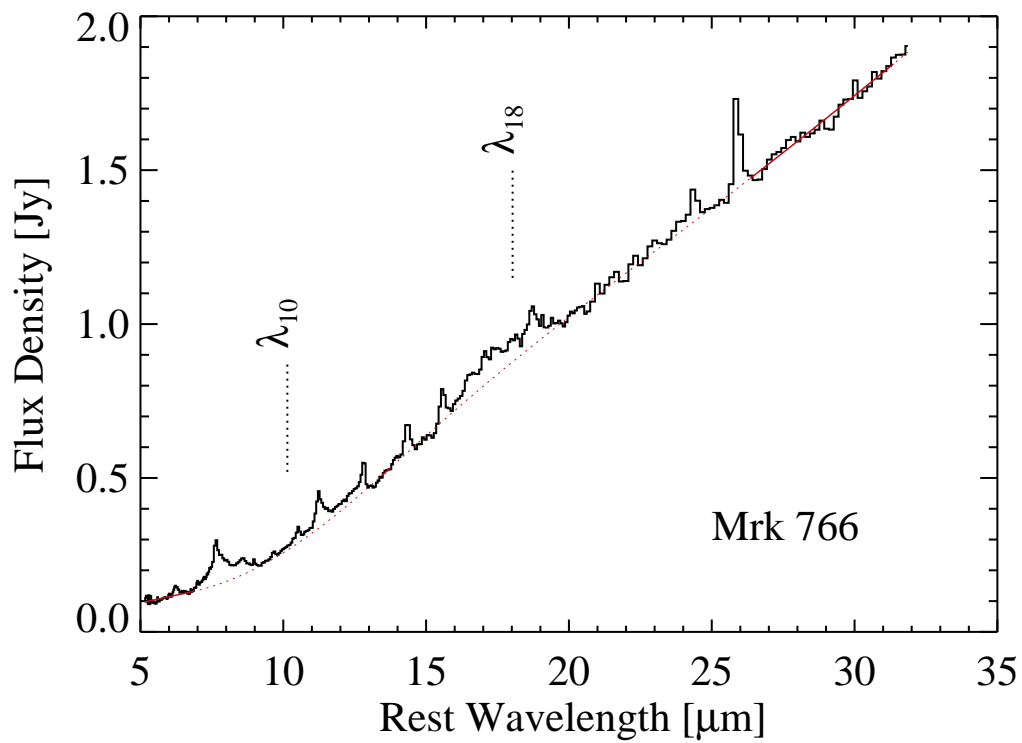}}
\caption{\label{fig3} Spectrum of the typical Seyfert 1 galaxy Mrk 766 illustrates the continuum fitting technique.  The resulting continuum (dotted red) is a spline fit to measurements over a short-wavelength region, an intermediate point around 14\um{}, and a long-wavelength region (solid red).  Vertical lines mark the measured peak wavelengths of the 10 and 18\um{} silicate features, which both appear in emission here, although the 10\um{} emission is very weak.}
\end{figure}

\clearpage
\begin{figure}[htb!]
\centerline{\includegraphics[width=6in]{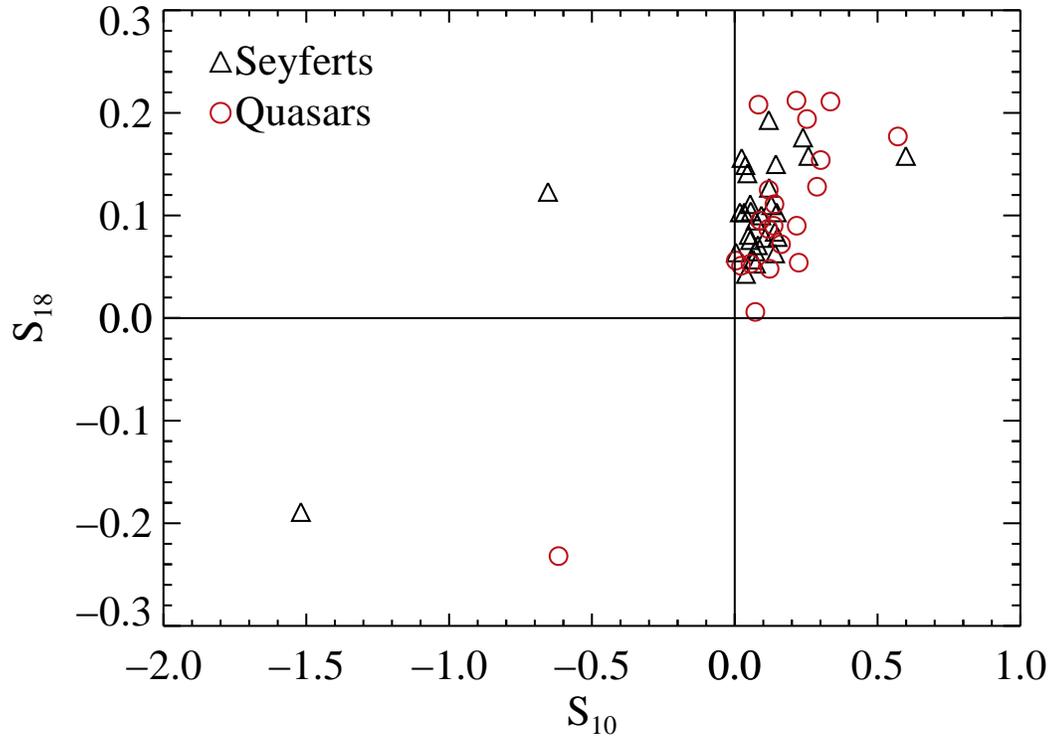}}
\caption{\label{fig4} Feature strengths at 10 and 18\um{} ($S_{10}$ and $S_{18}$).  Positive strengths show silicate emission in 49 of the 52 type 1 AGNs.  The only type 1 AGNs that do not show 10\um{} silicate emission are also LIRGs or ULIRGs.  Vertical and horizontal lines at zero strength separate the regions of emission ($S>0$) and absorption ($S<0$).  Triangles identify Seyfert 1 AGNs, and circles mark quasars.}
\end{figure}

\clearpage
\begin{figure}[htb!]
\centerline{\includegraphics[width=6in]{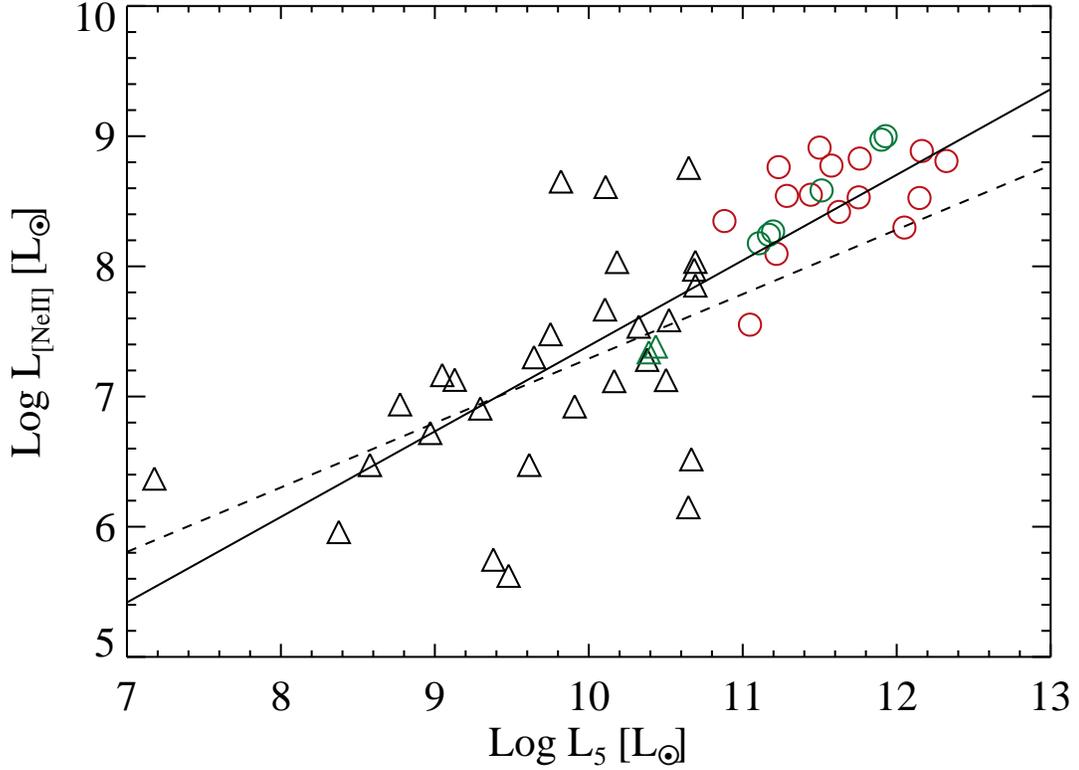}}
\caption{\label{fig5} AGN and star formation luminosities, which the 5\um{} monochromatic continuum luminosity and the [\ion{Ne}{2}] 12.8\um{} integrated luminosity indicate, are positively correlated, although star formation becomes proportionally less important with increasing AGN luminosity.  We certainly measure [\ion{Ne}{2}] in all but two Seyfert 1s and six quasars.  Considering the Seyferts and quasars separately and averaging spectra without detections, we estimate [\ion{Ne}{2}] in these cases (green).  The solid line shows the correlation over all galaxies, which agrees with the Seyfert 1 result using certain measurements alone (dashed).  Other symbols as in Figure~\ref{fig4}.}
\end{figure}

\clearpage
\begin{figure}[htb!]
\centerline{\includegraphics[width=6in]{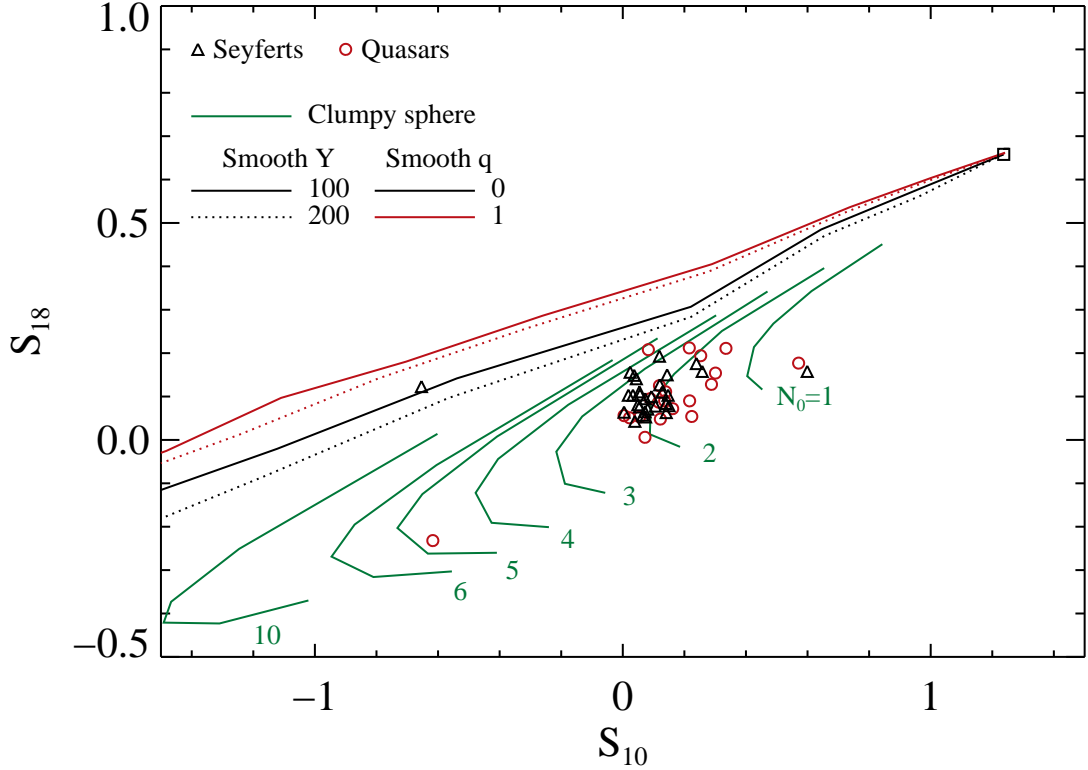}}
\caption{\label{fig6} Together, 10 and 18\um{} silicate feature strengths are sensitive to the dust geometry, and the AGN observations reveal clumpy surroundings.  Independent of geometry, all optically thin configurations are located at the same point in the diagram (square), and tracks of increasing optical depth move toward weaker emission and eventually show absorption.  All tracks of smooth spherical dust distributions (black and red), have similar slopes and lie separate from the data, for a range of spatial extent, $Y$, and radial density distribution ($\propto r^{-q}$).  Models of clumpy environments (green) occupy regions of this ``feature-feature diagram'' that are inaccessible to the smooth models.  Each clumpy sphere track is a function of the average number of clouds along radial rays, $N_{0}$, with $q=1$ and $Y=30$.  Clumpy models of $N_{0} \sim 2$ generally agree with the data.}
\end{figure}

\clearpage
\begin{figure}[htb!]
\centerline{\includegraphics[width=6in]{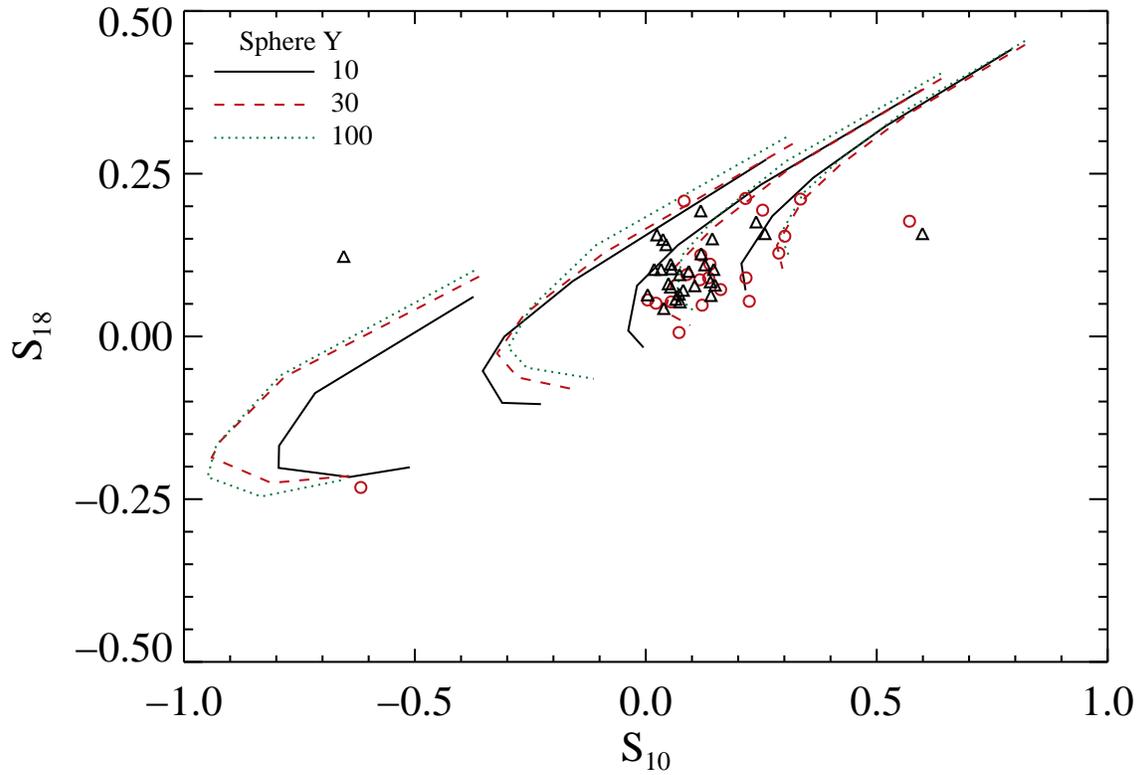}}
\caption{\label{fig7} Clumpy spherical distributions as a function of radial extent, $Y$, for $q=2$.  For $Y = 10$, 30, and 100, we plot curves of $N_{0}=1$, 2, 4, and 10.  Because the clouds closest to the AGN determine the MIR behavior, this inherently compact steep ($q=2$) distribution is insensitive to the outer radius.}
\end{figure}

\clearpage
\begin{figure}[htb!]
\centerline{\includegraphics[width=6in]{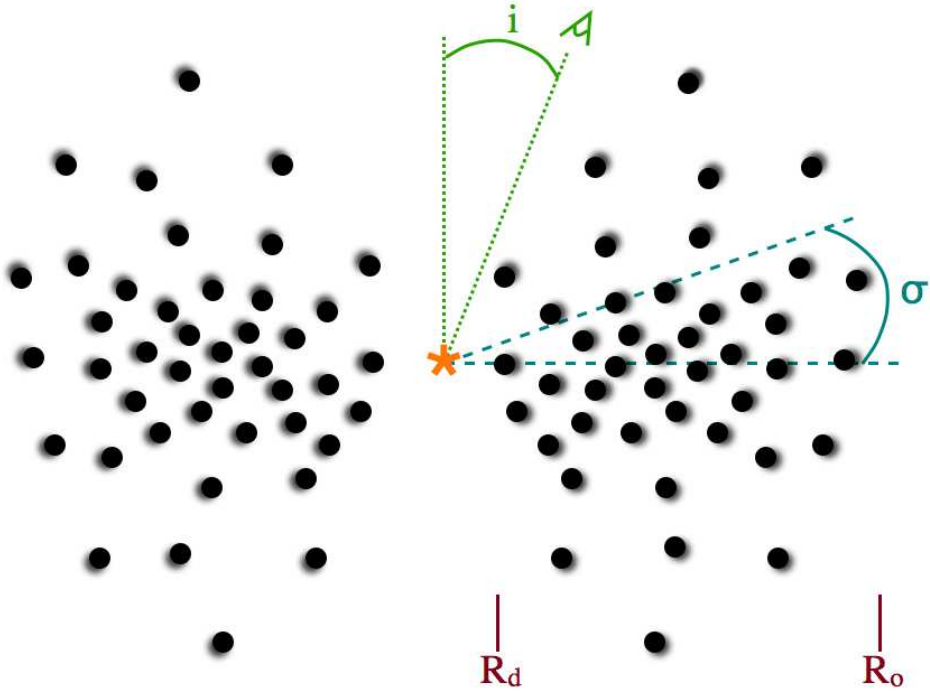}}
\caption{\label{fig8} Cartoon of the clumpy torus.  Emission and obscuration of the central engine are functions of viewing angle, $i$.  Clouds are distributed from the dust sublimation radius, $R_d$, to the outer radius, $R_o$, according to a radial power law.  The average number of clouds along an equatorial ray is $N_0$.  The scale height of the distribution is $\sigma$, with the average number of clouds $N_{los}(\beta) = N_0 \exp(-\beta^2/\sigma^2)$ along a radial ray at angle $\beta$ measured from the equator.}
\end{figure}

\clearpage
\begin{figure}[htb!]
\centerline{\includegraphics[width=6in]{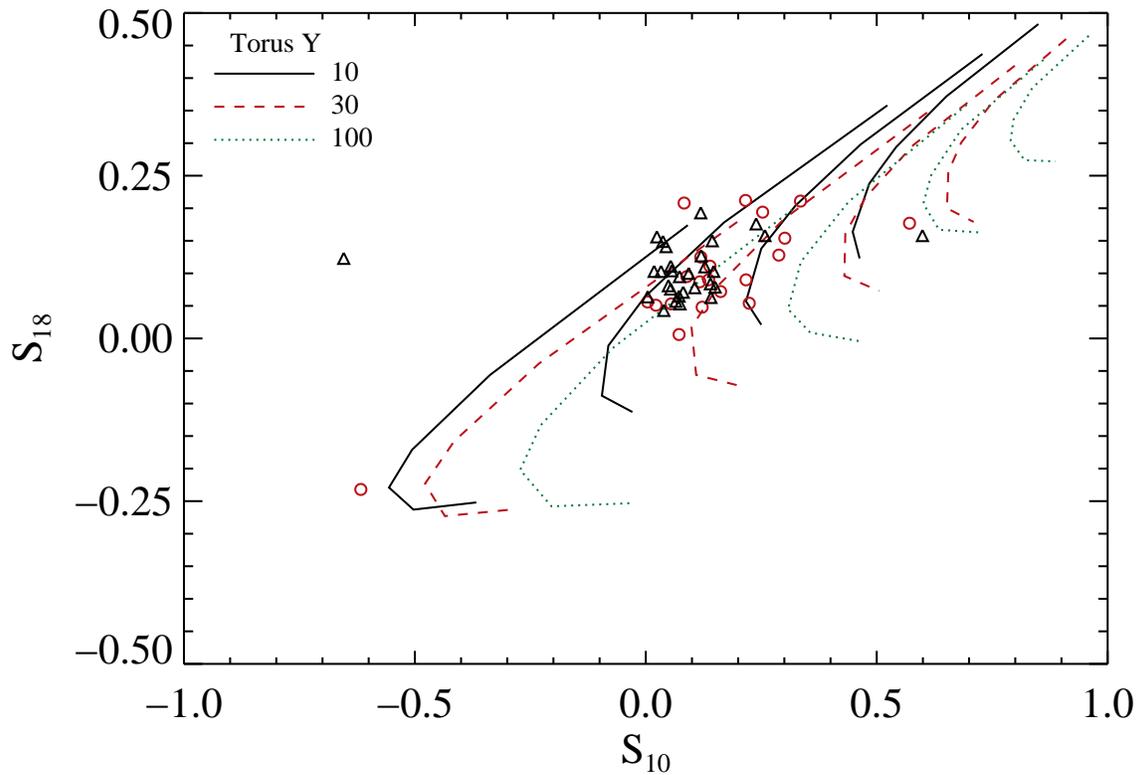}}
\caption{\label{fig9} Clumpy torus models as a function of $Y$, for $q=0$.  For $Y=10$, 30, and 100, we plot curves of $N_{0}=1$, 2, 4, and 10, fixing $\sigma=45^{\circ}$ and $i=30^{\circ}$.  In this distribution that is constant with radius, the size of the torus and $N_0$ are related, with increasing values of both $Y$ and $N_0$ together producing results similar to models having smaller size and cloud number.  The clouds close to the AGN govern the MIR emission, so in a large torus, $N_0$ must increase to provide enough clouds at small radius to match the features of a smaller torus having fewer clouds that are all confined to the small scale.}
\end{figure}

\clearpage
\begin{figure}[htb!]
\centerline{\includegraphics[width=6in]{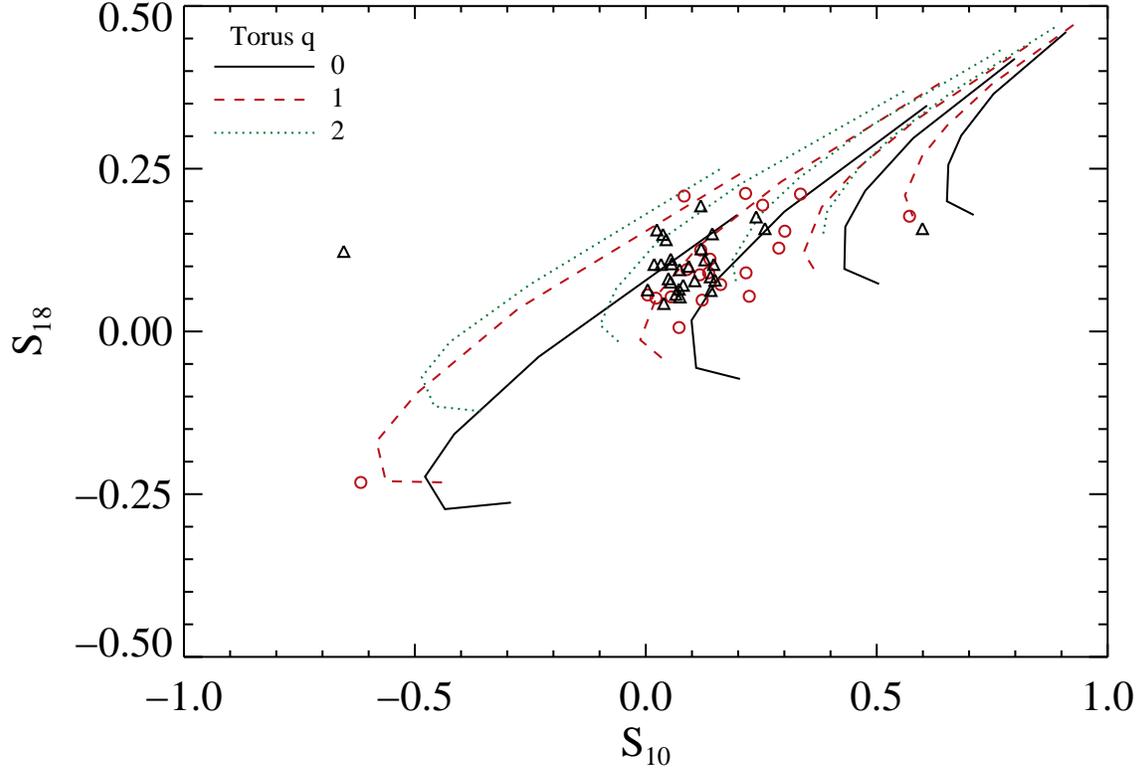}}
\caption{\label{fig10} Clumpy torus models as a function of radial density profile, $q$.  We plot curves for $N_{0}=1$, 2, 4, and 10, fixing $Y=30$ and $i=30^{\circ}$.  Increasing $q$ places more clouds close to the AGN for a given value of $N_0$.  Thus, the models having steeper density distributions require fewer clouds to match the silicate emission of the bulk of the data.  The shallowest density profile ($q=0$) does not follow the general trend of decreasing strength as rapidly as the steeper distributions with increasing $N_0$.  A compact torus (effective size $\sim 15R_d$) having few clouds along radial rays ($\lesssim 6$ along the equator) accounts for the MIR silicate features of unobscured AGNs.}
\end{figure}

\clearpage
\begin{figure}[htb!]
\centerline{\includegraphics[width=6in]{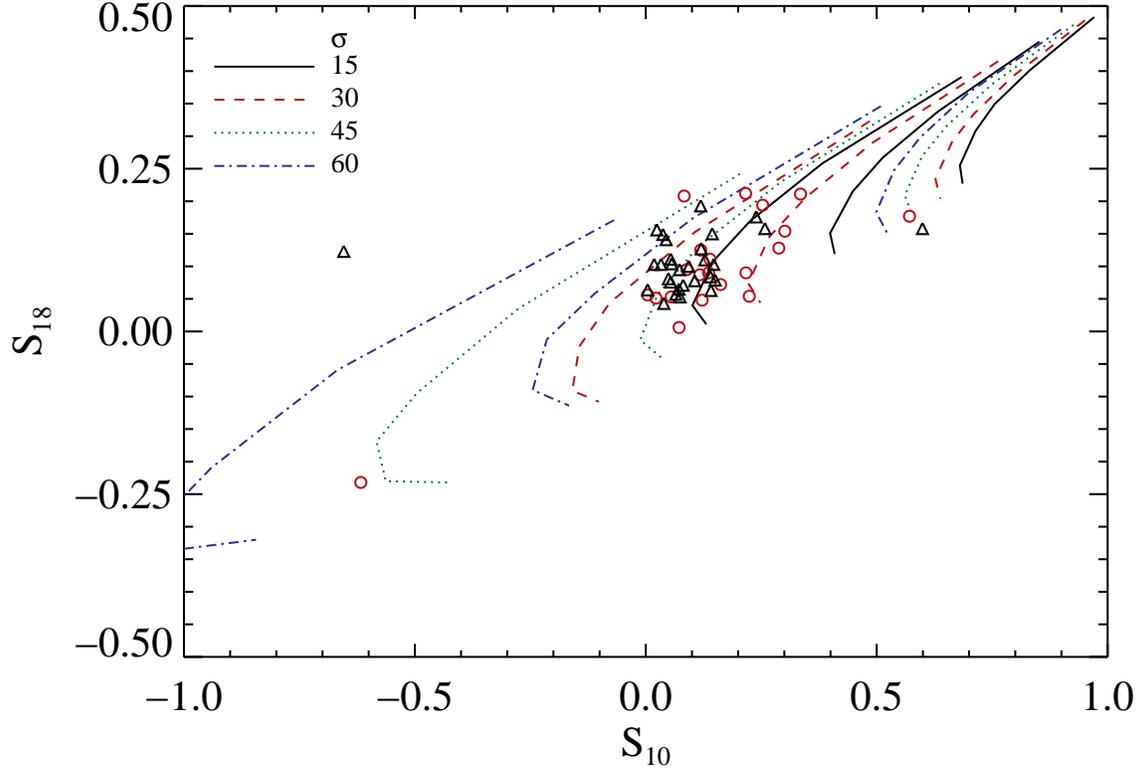}}
\caption{\label{fig11} Clumpy torus models as a function of $\sigma$.  We plot curves of $N_0=1$, 4, and 10, fixing $Y=30$, $q=1$, and $i=30^{\circ}$.  Increasing the torus scale height reduces the silicate strength for a given $N_0$, obeying the general trend of decreasing strength with increasing total cloud number.  However, even for $N_{0}=10$, the $\sigma=15^{\circ}$ model shows emission, because this thin torus contains fewer clouds to block direct views of bright cloud faces.  We measure the same effect for other values of $Y$ and $q$.}
\end{figure}

\clearpage
\begin{figure}[htb!]
\centerline{\includegraphics[width=6in]{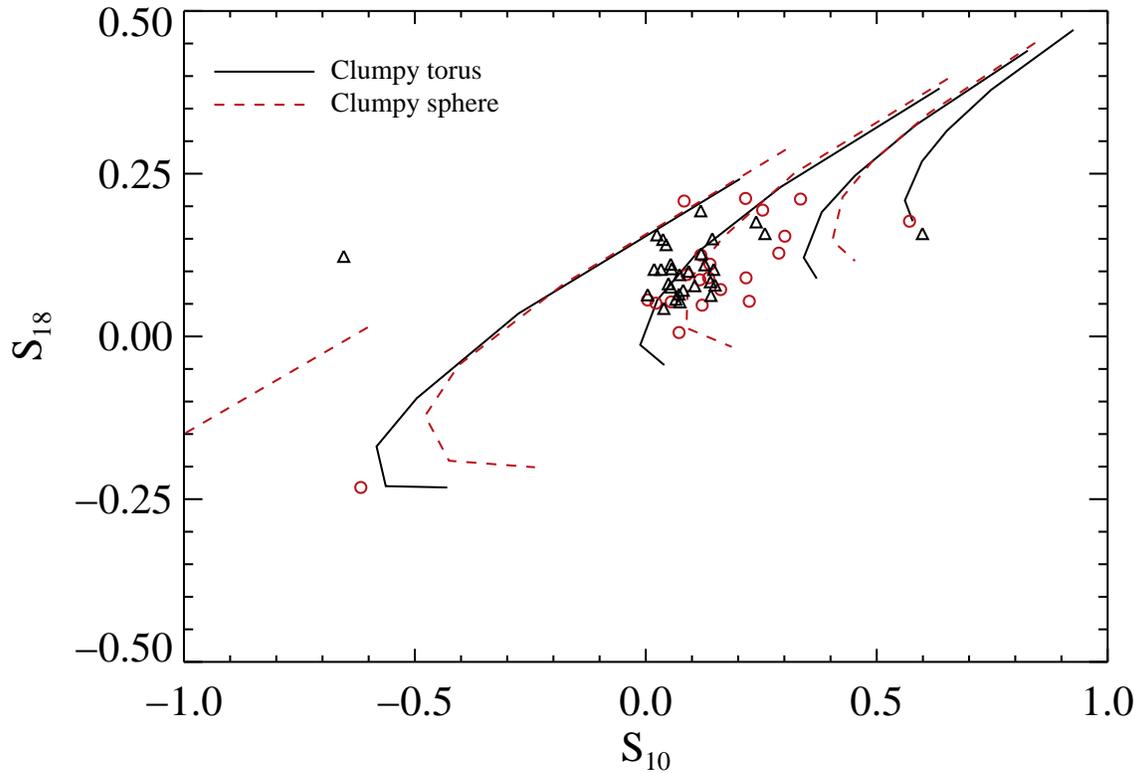}}
\caption{\label{fig12} Clumpy torus and clumpy sphere models.  We plot curves of $N_0= 1$, 2, 4, and 10, fixing $Y=30$ and $q=1$ in both cases, and i=$30^{\circ}$ and $\sigma=45^{\circ}$ in the torus models.  The sphere and torus results are extremely similar for different values of $N_0$, with larger $N_0$ in the torus.  Because all clouds reprocess the AGN light and contribute to the MIR emission, the larger value of the parameter $N_0$ in the toroidal distribution is required to have the same total number of clouds as the spherical distribution with smaller $N_0$.}
\end{figure}
\clearpage

\end{document}

%% file: tab1.tex
\begin{deluxetable}{lrrrrrrrr}
\tablewidth{0pt}
\tablecolumns{9}
\tabletypesize{\tiny}
\tablecaption{Observations and Galaxy Data\label{tab:sources}}
\tablehead{
&&&&&& &\multicolumn{2}{c}{On-Source Exposure}\\
\cline{8-9}\\[-2ex]
   &\colhead{R.A.}      &\colhead{Decl.}   &  &\colhead{Scale}   & & &\colhead{ShortLow}  &\colhead{LongLow}\\  
\colhead{Galaxy} &   \colhead{(J2000.0)}  &\colhead{(J2000.0)}  & \colhead{$z$}  &\colhead{(pc arcsec$^{-1}$)}  & \colhead{AOR} & \colhead{Observation Date} &\colhead{(s)}&\colhead{(s)}
}
\startdata
\sidehead{\underline{Seyfert 1s}}
ESO 12-G21         &  00 40 46.2      &   $-$79 14 24      & 0.030      & 610               & 12465920         &  2005-APR-17          &   24      &   24  \\
Mrk 335            &  01 22 40.8      &     +26 52 06      & 0.026      & 530               & 12450560         &  2004-DEC-13          &   24      &   24  \\
Mrk 1034NE         &  02 23 20.4      &     +32 11 34      & 0.034      & 700               & 20320512         &  2007-SEP-01          &  480      &  160  \\ 
NGC 931            &  02 28 14.5      &     +31 18 42      & 0.017      & 340               & 12460032         &  2005-JAN-15          &   24      &   24  \\
IRAS F03450+0055   &  03 47 40.2      &     +01 05 14      & 0.031      & 630               & 4674816          &  2004-AUG-31          &   24      &   56  \\
NGC 1566           &  04 20 00.4      &   $-$54 56 16      & 0.005      & 100               & 9490688          &  2004-JUL-18          &   56      &   60  \\
3C120              &  04 33 11.1      &     +05 21 16      & 0.033      & 670               & 4847360          &  2004-OCT-03          &  112      &  120  \\
MCG -5-13-17       &  05 19 35.8      &   $-$32 39 28      & 0.012      & 260               & 12468480         &  2005-MAR-14          &   24      &   24  \\
Mrk 6              &  06 52 12.3      &     +74 25 38      & 0.019      & 390               & 12483584         &  2004-NOV-17          &   24      &   24  \\
Mrk 9              &  07 36 57.0      &     +58 46 13      & 0.040      & 800               & 12483072         &  2005-MAR-11          &   24      &   24  \\
Mrk 79             &  07 42 32.8      &     +49 48 35      & 0.022      & 450               & 12453632         &  2005-MAR-18          &   24      &   24  \\
Mrk 704            &  09 18 26.0      &     +16 18 19      & 0.029      & 590               & 12444416         &  2005-APR-18          &   24      &   24  \\
UGC 5101           &  09 35 51.7      &     +61 21 11      & 0.039      & 790               & 4973056          &  2004-MAR-23          &  168      &  240  \\
Mrk 1239           &  09 52 19.1      &   $-$01 36 44      & 0.020      & 410               & 12453120         &  2005-MAY-26          &   24      &   24  \\
NGC 3227           &  10 23 30.6      &     +19 51 54      & 0.004      & 80                & 4934656          &  2005-DEC-10          &   56      &   56  \\
NGC 3511           &  11 03 23.8      &   $-$23 05 12      & 0.004      & 77                & 12473600         &  2005-MAY-25          &   24      &   24  \\
NGC 3516           &  11 06 47.5      &     +72 34 07      & 0.009      & 180               & 12473344         &  2004-DEC-12          &   24      &   24  \\
NGC 4051           &  12 03 09.6      &     +44 31 53      & 0.002      & 48                & 12451072         &  2005-JAN-14          &   24      &   24  \\
NGC 4151           &  12 10 32.6      &     +39 24 21      & 0.003      & 69                & 3754496          &  2004-JAN-08          &  112      &   96  \\
Mrk 766            &  12 18 26.5      &     +29 48 46      & 0.013      & 270               & 12465408         &  2005-MAY-22          &   24      &   24  \\
NGC 4593           &  12 39 39.4      &   $-$05 20 39      & 0.009      & 190               & 12457216         &  2005-JUL-01          &   24      &   24  \\
MCG -2-33-34       &  12 52 12.5      &   $-$13 24 53      & 0.017      & 340               & 12481280         &  2005-FEB-16          &   24      &   24  \\
MCG -6-30-15       &  13 35 53.8      &   $-$34 17 44      & 0.008      & 160               & 4849920          &  2005-FEB-15          &  112      &  120  \\
IC 4329A           &  13 49 19.2      &   $-$30 18 34      & 0.016      & 330               & 4848640          &  2004-JUL-13          &  112      &  120  \\
NGC 5548           &  14 17 59.5      &     +25 08 12      & 0.017      & 350               & 4855296          &  2004-JUL-14          &  112      &  120  \\
Mrk 817            &  14 36 22.1      &     +58 47 39      & 0.032      & 640               & 12461056         &  2004-DEC-14          &   24      &   24  \\
NGC 6860           &  20 08 46.9      &   $-$61 06 01      & 0.015      & 310               & 12462592         &  2005-APR-24          &   24      &   24  \\
Mrk 509            &  20 44 09.7      &   $-$10 43 25      & 0.034      & 700               & 4850432          &  2004-MAY-14          &  112      &  120  \\
NGC 7213           &  22 09 16.3      &   $-$47 10 00      & 0.006      & 120               & 4856320          &  2004-MAY-15          &  112      &  120  \\
NGC 7469           &  23 03 15.6      &     +08 52 26      & 0.016      & 330               & 3755008          &  2003-DEC-16          &  112      &  120  \\
NGC 7603           &  23 18 56.6      &     +00 14 38      & 0.030      & 600               & 10870784         &  2004-DEC-08          &  280      &  280  \\
\sidehead{\underline{Quasars}}
PG0052+251         &  00 54 52.1      &     +25 25 38     & 0.155      & 3500              & 4675072           & 2004-JAN-04           &  56       &  120  \\
3C048              &  01 37 41.3      &     +33 09 35     & 0.367      & 9300              & 4670720           & 2004-AUG-07           &  24       &   56  \\  
IRAS F07599+6508   &  08 04 33.1      &     +64 59 49     & 0.148      & 3300              & 17103104          & 2006-APR-22           & 168       &  240  \\
PG0947+396         &  09 50 48.4      &     +39 26 51     & 0.206      & 4800              & 14190592          & 2005-DEC-13           & 480       &  720  \\        
PG0953+414         &  09 56 52.4      &     +41 15 22     & 0.234      & 5500              & 4675328           & 2004-APR-17           &  56       &  120  \\
3C234              &  10 01 49.6      &     +28 47 09     & 0.185      & 4200              & 11305728          & 2005-APR-22           & 480       &  720  \\    
PG1048+342         &  10 51 43.9      &     +33 59 27     & 0.167      & 3800              & 14192128          & 2006-MAY-27           & 480       &  720  \\      
PG1116+215         &  11 19 08.6      &     +21 19 18     & 0.177      & 4000              & 4734464           & 2004-MAY-14           &  24       &   56  \\ 
PG1121+422         &  11 24 39.2      &     +42 01 45     & 0.225      & 5300              & 14193664          & 2005-DEC-19           & 480       &  720  \\          
3C273              &  12 29 06.7      &     +02 03 09     & 0.158      & 3600              & 4978176           & 2004-JAN-06           & 168       &  168  \\
Mrk 231            &  12 56 14.2      &     +56 52 25     & 0.042      & 880               & 4978688           & 2004-APR-14           & 112       &  120  \\
PG1307+085         &  13 09 47.0      &     +08 19 49     & 0.155      & 3500              & 4735488           & 2006-JAN-21           &  24       &   56  \\
PG1309+355         &  13 12 17.8      &     +35 15 21     & 0.184      & 4200              & 4736000           & 2004-MAY-14           &  24       &   56  \\      
PG1322+659         &  13 23 49.5      &     +65 41 48     & 0.168      & 3800              & 14196224          & 2006-JUN-21           & 480       &  720  \\        
IRAS F13349+2438   &  13 37 18.7      &     +24 23 03     & 0.108      & 2400              & 4373760           & 2005-JUN-07           & 280       &  280  \\
PG1352+183         &  13 54 35.6      &     +18 05 17     & 0.152      & 3400              & 4736512           & 2006-JAN-21           &  24       &   56  \\       
PG1354+213         &  13 56 32.7      &     +21 03 52     & 0.300      & 7400              & 14196992          & 2006-FEB-02           & 480       &  720  \\       
PG1402+261         &  14 05 16.2      &     +25 55 35     & 0.164      & 3700              & 4675584           & 2004-JUN-08           &  56       &  120  \\         
PG1427+480         &  14 29 43.1      &     +47 47 26     & 0.221      & 5200              & 14198528          & 2006-JAN-18           & 480       &  720  \\              
PG2130+099         &  21 32 27.8      &     +10 08 20     & 0.063      & 1400              & 3761408           & 2004-JUN-06           & 168       &  240  \\  
PG2233+134         &  22 36 07.7      &     +13 43 55     & 0.326      & 8100              & 4734208           & 2003-DEC-17           &  56       &  120  \\ 
																	         		    		       
\enddata																         		    
\tablecomments{Units of right ascension are hours, minutes, and seconds, and units of declination are degrees, arcminutes, and arcseconds.}      		    
																	         		       
\end{deluxetable}

%% file: tab2.tex
\begin{deluxetable}{ccc}
\tablewidth{0pt}
\tabletypesize{\tiny}
\tablecaption{Average Type 1 AGN Spectra\label{table:avgspectra}}
\tablehead{\colhead{Rest Wavelength}  &\colhead{Seyfert 1 Scaled Flux Density}      &\colhead{Quasar Scaled Flux Density}\\
	   \colhead{(\um)}            &\colhead{(Jy)}                               &\colhead{(Jy)}\\
           \colhead{(1)}              &\colhead{(2)}                                &\colhead{(3)}}

\startdata

5.000   & 0.226   & 0.441 \\
5.030	& 0.237   & 0.446 \\
5.060	& 0.247   & 0.449 \\
5.090	& 0.257   & 0.453 \\
5.120	& 0.266   & 0.450 \\
5.150	& 0.271   & 0.449 \\
5.180	& 0.277   & 0.453 \\
5.210	& 0.287   & 0.458 \\

\enddata																         		    

\tablecomments{Table \ref{table:avgspectra} is published in its entirety in the electronic edition of the {\it Astrophysical Journal}.  A portion is shown here for guidance regarding its form and content.  Col. (1): Rest wavelength. The wavelength scale is non-uniform and based on the orginal resolution, which decreases toward longer wavelengths.  Col. (2): Average Seyfert 1 spectrum normalized at 14\um.  Col. (3): Average quasar spectrum normalized at 14\um.}      		    
																	         		       
\end{deluxetable}															         		    		       
	

%% file: tab3.tex
\begin{deluxetable}{lrrrrccrrr}
\tablewidth{0pt}
\tablecolumns{10}
\tabletypesize{\tiny}
\tablecaption{Spectral Measurements\label{tab:observed}}
\tablehead{
                  &                                   &                                    &                              &                                      &\multicolumn{2}{c}{Integrated Flux}\\
\cline{6-7}\\[-2ex] 
  &       &\colhead{$\lambda_{10}$}  &  &\colhead{$\lambda_{18}$}    &\colhead{[Ne II]}                                                  &\colhead{6.2\um{} PAH}   &\colhead{$F_{5}$}         &\colhead{$F_{14}$}  &\colhead{$F_{30}$}\\
\colhead{Galaxy} &  \colhead{$S_{10}$}  &\colhead{(\um)}           &  \colhead{$S_{18}$}  &\colhead{(\um)}  &\colhead{(erg s$^{-1}$ cm$^{-2}$)}                                           &\colhead{(erg s$^{-1}$ cm$^{-2}$)} &\colhead{(Jy)}                      &\colhead{(Jy)}                &\colhead{(Jy)}}

\startdata
\sidehead{\underline{Seyfert 1s}}
ESO 12-G21         &  0.15    &  10.1   &  0.08   &  18.2   &  2.1E-13   &   5.9E-13  &  0.05  &  0.15   &  0.34\\
Mrk 335            &  0.15    &  10.0   &  0.10   &  18.3   &  \nodata   &   \nodata  &  0.11  &  0.22   &  0.35\\
Mrk 1034           & $-$0.65  &  10.0   &  0.12   &  18.3   &  6.9E-13   &   1.5E-12  &  0.02  &  0.17   &  1.00\\ 
NGC 931            &  0.05    &  10.1   &  0.08   &  18.2   &  8.6E-14   &   9.6E-14  &  0.16  &  0.54   &  1.10\\
IRAS F03450+0055   &  0.14    &   9.9   &  0.06   &  18.0   &  \nodata   &   \nodata  &  0.09  &  0.31   &  0.49\\
NGC 1566           &  0.02    &  10.0   &  0.16   &  18.2   &  2.2E-13   &   7.5E-13  &  0.05  &  0.15   &  0.43\\
3C120              &  0.26    &  10.2   &  0.15   &  17.8   &  6.3E-14   &   1.0E-13  &  0.09  &  0.31   &  0.63\\
MCG -5-13-17       &  0.09    &   9.8   &  0.10   &  18.2   &  1.6E-13   &   2.3E-13  &  0.03  &  0.21   &  0.64\\
Mrk 6              &  0.24    &  10.2   &  0.18   &  18.1   &  2.4E-13   &   2.0E-13  &  0.11  &  0.32   &  0.70\\
Mrk 9              &  0.04    &   9.7   &  0.14   &  17.8   &  7.8E-14   &   \nodata  &  0.09  &  0.24   &  0.52\\
Mrk 79             &  0.07    &  10.1   &  0.10   &  18.0   &  7.0E-14   &   \nodata  &  0.15  &  0.48   &  1.07\\
Mrk 704            &  0.07    &  10.2   &  0.05   &  18.2   &  2.9E-15   &   \nodata  &  0.16  &  0.44   &  0.47\\
UGC 5101           & $-$1.52  &   9.9   & $-$0.19 &  18.3   &  6.4E-13   &   1.4E-13  &  0.09  &  0.27   &  2.24\\
Mrk 1239           &  0.13    &  10.1   &  0.11   &  18.3   &  1.5E-14   &   2.5E-13  &  0.36  &  0.82   &  1.37\\
NGC 3227           &  0.01    &  10.0   &  0.06   &  18.2   &  1.1E-12   &   1.6E-12  &  0.13  &  0.63   &  1.91\\
NGC 3511           &  0.07    &  10.2   &  0.07   &  18.2   &  3.2E-13   &   3.8E-13  &  0.004  &  0.06   &  0.22\\
NGC 3516           &  0.03    &  10.0   &  0.10   &  17.7   &  9.9E-15   &   9.3E-14  &  0.12  &  0.39   &  0.92\\
NGC 4051           &  0.07    &  10.0   &  0.06   &  18.0   &  3.1E-13   &   7.2E-13  &  0.14  &  0.64   &  1.43\\
NGC 4151           &  0.14    &  10.0   &  0.15   &  18.4   &  1.3E-12   &   6.2E-13  &  0.56  &  2.45   &  4.06\\
Mrk 766            &  0.05    &  10.1   &  0.08   &  18.0   &  3.3E-13   &   3.5E-13  &  0.11  &  0.57   &  1.75\\
NGC 4593           &  0.11    &  10.0   &  0.08   &  18.1   &  6.8E-14   &   1.6E-13  &  0.16  &  0.46   &  0.97\\
MCG -2-33-34       &  0.04    &   9.8   &  0.15   &  17.8   &  1.2E-13   &   2.1E-13  &  0.02  &  0.12   &  0.39\\
MCG -6-30-15       &  0.02    &  10.2   &  0.10   &  18.3   &  1.7E-14   &   1.0E-13  &  0.13  &  0.45   &  0.79\\
IC 4329A           &  0.04    &   9.9   &  0.04   &  18.1   &  2.4E-13   &   \nodata  &  0.25  &  1.36   &  2.05\\
NGC 5548           &  0.14    &  10.1   &  0.08   &  18.1   &  1.2E-13   &   7.9E-14  &  0.05  &  0.29   &  0.59\\
Mrk 817            &  0.08    &  10.1   &  0.07   &  18.0   &  2.4E-14   &   1.5E-13  &  0.10  &  0.40   &  1.36\\
NGC 6860           &  0.06    &  10.1   &  0.10   &  18.0   &  6.9E-14   &   \nodata  &  0.11  &  0.25   &  0.36\\
Mrk 509            &  0.12    &  10.1   &  0.19   &  18.0   &  1.6E-13   &   3.0E-13  &  0.13  &  0.35   &  0.66\\
NGC 7213           &  0.60    &  10.1   &  0.16   &  17.9   &  2.8E-13   &   7.0E-14  &  0.09  &  0.28   &  0.46\\
NGC 7469           &  0.05    &   9.8   &  0.11   &  17.8   &  2.8E-12   &   4.6E-12  &  0.15  &  1.41   &  7.96\\
NGC 7603           &  0.12    &  10.2   &  0.13   &  17.7   &  1.9E-13   &   4.7E-13  &  0.17  &  0.27   &  0.33\\
\sidehead{\underline{Quasars}}				  
PG0052+251         &  0.30    &  10.2   &  0.15   &  17.9   &  2.1E-14   &   7.6E-14  &  0.02  &  0.06   &  0.05\\
3C048              &  0.16    &  10.3   &  0.07   &  18.3   &  6.7E-15   &   \nodata  &  0.02  &  0.10   &  0.39\\ 
IRAS F07599+6508   &  0.07    &  10.0   &  0.01   &  17.8   &  2.3E-14   &   8.6E-14  &  0.16  &  0.30   &  0.84\\
PG0947+396         &  0.09    &  10.3   &  0.10   &  17.9   &  1.2E-14   &   \nodata  &  0.02  &  0.03   &  0.05\\
PG0953+414         &  0.25    &   9.8   &  0.19   &  17.9   &  1.7E-14   &   1.1E-13  &  0.02  &  0.04   &  0.04\\
3C234              &  0.01    &  10.0   &  0.06   &  17.9   &  1.4E-14   &   1.2E-15  &  0.04  &  0.20   &  0.27\\
PG1048+342         &  0.22    &  10.2   &  0.21   &  18.4   &  1.2E-14   &   \nodata  &  0.01  &  0.02   &  0.02\\ 
PG1116+215         &  0.22    &  10.2   &  0.09   &  18.2   &  \nodata   &   \nodata  &  0.06  &  0.09   &  0.11\\ 
PG1121+422         &  0.12    &  10.0   &  0.09   &  18.2   &  \nodata   &   \nodata  &  0.01  &  0.01   &  0.01\\ 
3C273              &  0.12    &  10.1   &  0.05   &  18.1   &  3.8E-14   &   3.2E-14  &  0.21  &  0.38   &  0.55\\
Mrk 231            & $-$0.62  &   9.8   & $-$0.23 &  17.9   &  5.8E-13   &   5.9E-13  &  0.63  &  2.98   &  13.17\\
PG1307+085         &  0.34    &   9.8   &  0.21   &  17.7   &  \nodata   &   \nodata  &  0.01  &  0.05   &  0.06\\
PG1309+355         &  0.57    &  10.0   &  0.18   &  17.7   &  3.5E-14   &   \nodata  &  0.02  &  0.07   &  0.10\\ 
PG1322+659         &  0.14    &  10.0   &  0.11   &  18.2   &  6.5E-15   &   1.7E-14  &  0.01  &  0.03   &  0.05\\ 
IRAS F13349+2438   &  0.06    &  10.1   &  0.05   &  17.8   &  2.7E-14   &   \nodata  &  0.26  &  0.56   &  0.72\\
PG1352+183         &  0.08    &  10.3   &  0.21   &  18.2   &  \nodata   &   \nodata  &  0.02  &  0.02   &  0.03\\ 
PG1354+213         &  0.12    &   9.9   &  0.13   &  17.9   &  \nodata   &   \nodata  &  0.01  &  0.02   &  0.04\\ 
PG1402+261         &  0.22    &  10.0   &  0.05   &  17.9   &  1.4E-14   &   \nodata  &  0.04  &  0.08   &  0.16\\ 
PG1427+480         &  0.14    &  10.3   &  0.09   &  17.7   &  1.6E-14   &   \nodata  &  0.01  &  0.02   &  0.07\\ 
PG2130+099         &  0.02    &  10.0   &  0.05   &  18.3   &  1.5E-14   &   4.3E-14  &  0.08  &  0.22   &  0.35\\
PG2233+134         &  0.29    &   9.8   &  0.13   &  17.7   &  \nodata   &   \nodata  &  0.02  &  0.04   &  0.07\\ 
		      
\enddata	      	
\end{deluxetable}

%% file: ms.bbl
\begin{thebibliography}{}

\bibitem[Alonso-Herrero et al.(2006)]{alonso} Alonso-Herrero, A., et al. 2006, \apj, 640, 167

\bibitem[Alonso-Herrero et al.(2006)]{alo06} Alonso-Herrero, A., Colina, L., Packham, C., D{\'{\i}}az-Santos, T., Rieke, G. H., Radomski, J. T., \& Telesco, C. M.\ 2006, \apjl, 652, L83

\bibitem[Antonucci(1993)]{antonucci} Antonucci, R. 1993, \araa, 31, 473

\bibitem[Beckert et al.(2008)]{beckert08} Beckert, T., Driebe, T., H\"{o}nig, S. F., \& Weigelt, G. 2008, \aap, 486, 17

\bibitem[Buchanan et al.(2006)]{buchanan06} Buchanan, C. L., Gallimore, J. F., O'Dea, C. P., Baum, S. A., Axon, D. J., Robinson, A., Elitzur, M., \& Elvis, M. 2006, \aj, 132, 401

\bibitem[Draine(2003a)]{drainea} Draine, B. T. 2003a, \apj, 598, 1017

\bibitem[Draine(2003b)]{draineb} Draine, B. T. 2003b, \apj, 598, 1026

\bibitem[Efstathiou \& Rowan-Robinson(1995)]{Row95} Efstathiou, A., \& Rowan-Robinson, M. 1995, \mnras, 273, 649

\bibitem[Elvis et al.(1994)]{elvis} Elvis, M., et al. 1994, \apjs, 95, 1

\bibitem[Elvis et al.(2004)]{Elv04} Elvis, M., Risaliti, G., 
Nicastro, F., Miller, J.~M., Fiore, F., \& Puccetti, S.\ 2004, \apjl, 615, L25 

\bibitem[Genzel et al.(1998)]{Gen98} Genzel, R., et al. 1998, \apj, 498, 579 

\bibitem[Granato \& Danese(1994)]{granato94} Granato, G. L., \& Danese, L. 1994, \mnras, 268, 235

\bibitem[Hao et al.(2005)]{hao05} Hao, L., et al. 2005, \apj, 625, 75

\bibitem[Hao et al.(2007)]{hao07} Hao, L., Weedman, D. W., Spoon, H. W. W., Marshall, J. A., Levenson, N. A., Elitzur, M., \& Houck, J. R. 2007, \apj, 655, 77

\bibitem[Ho \& Keto(2007)]{hoketo} Ho, L. C., \& Keto, E. 2007, \apj, 658, 314

\bibitem[H\"{o}nig et al.(2006)]{honig06} H\"{o}nig, S. F., Beckert, T., Ohnaka, K., \& Weigelt, G. 2006, \aap, 452, 459

\bibitem[H\"{o}nig et al.(2008)]{honig08} H\"{o}nig, S. F., Smette, A., Beckert, T., Horst, H., Duschl, W., Gandhi, P., Kishimoto, M., \& Weigelt, G. 2008, \aap, 485, 21

\bibitem[Horst et al.(2006)]{horst06} Horst, H., Smette, A., Gandhi, P., \& Duschl, W. J. 2006, \aap, 457, 17

\bibitem[Houck et al.(2004)]{houck} Houck, J. R., et al. 2004, \apjs, 154, 18

\bibitem[Jaffe et al.(2004)]{jaffe04} Jaffe, W., et al. 2004, \nat, 429, 47

\bibitem[Keremedjiev et al.(2008)]{Keremedjiev08} Keremedjiev, M., Hao, L., \& Charmandaris, V. 2008, arXiv:0806.2910.

\bibitem[Klaas et al.(2001)]{klaas} Klaas, U., et al. 2001, \aap, 379, 823

\bibitem[Knacke \& Thomson(1973)]{knacke73} Knacke, R. F., \& Thomson, R. K. 1973, \pasp, 85, 341

\bibitem[Krolik \& Begelman(1988)]{krolikbegelman88} Krolik, J. H., \& Begelman, M. C. 1988, \apj, 329, 702

\bibitem[Levenson et al.(2007)]{lev} Levenson, N. A., Sirocky, M. M., Hao, L., Spoon, H. W. W., Marshal, J. A., Elitzur, M., \& Houck, J. R. 2007, \apj, 654, 45

\bibitem[Lutz et al.(2003)]{lutz03} Lutz, D., Sturm, E., Genzel, R., Spoon, H. W. W., Moorwood, A. F. M., Netzer, H., \& Sternberg, A. 2003, \aap, 409, 867

\bibitem[Lutz et al.(2004)]{lutz04} Lutz, D., Maiolino, R., Spoon, H. W. W., \& Moorwood, A. F. M. 2004, \aap, 418, 465

\bibitem[Mason et al.(2006)]{mason06} Mason, R. E., Geballe, T. R., Packham, C., Levenson, N. A., Elitzur, M., Fisher, R. S., \& Perlman, E. 2006, \apj, 640, 612

\bibitem[Mel\'{e}ndez et al.(2008)]{Melendez} Mel\'{e}ndez, M., et al. 2008, \apj, 682, 94

\bibitem[Nenkova et al.(2002)]{nenkova02} Nenkova, M., Ivezi\'{c}, \u{Z}., \& Elitzur, M. 2002, \apj, 570, 9

\bibitem[Nenkova et al.(2008a)]{Nenkova08a} Nenkova, M., Sirocky, M. M., Ivezi\'{c}, \u{Z}., \& Elitzur, M. 2008a, \apj, 685, 147

\bibitem[Nenkova et al.(2008b)]{nenkova08} Nenkova, M., Sirocky, M. M., Nikutta, R., Ivezi\'{c}, \u{Z}., \& Elitzur, M. 2008b, 685, 160

\bibitem[Netzer et al.(2007)]{netzer07} Netzer, H., et al. 2007, \apj, 666,806

\bibitem[Ossenkopf et al.(1992)]{ossenkopf} Ossenkopf, V., Henning, T., \& Mathis, J. S. 1992, \aap, 261, 567

\bibitem[Pier \& Krolik(1992)]{pierkrolik92} Pier, E. A., \& Krolik, J. H.\ 1992, \apj, 401, 99

\bibitem[Risaliti et al.(2007)]{Ris07} Risaliti, G., Elvis, 
M., Fabbiano, G., Baldi, A., Zezas, A., \& Salvati, M.\ 2007, \apjl, 659, L111 

\bibitem[Rush et al.(1993)]{rush} Rush, B., Malkan, M. A., \& Spinoglio, L.\ 1993, \apjs, 89, 1

\bibitem[Sanders \& Mirabel(1996)]{sanders96} Sanders, D. B., \& Mirabel, I. F. 1996, \araa, 34, 749

\bibitem[Schartmann et al.(2008)]{schartmann08} Schartmann, M., Meisenheimer, K., Camenzind, M., Wolf, S., Tristram, K. R. W., \& Henning, T. 2008, \aap, 482, 67

\bibitem[Schmitt et al.(2001)]{Schmitt01} Schmitt, H. R., Antonucci, R. R. J., Ulvestad, J. S., Kinney, A. L., Clarke, C. J., \& Pringle, J. E. 2001, \apj, 555, 663

\bibitem[Schweitzer et al.(2006)]{schweitzer06} Schweitzer, M., et al. 2006, \apj, 649, 79

\bibitem[Schweitzer et al.(2008)]{schweitzer08} Schweitzer, M., et al. 2008, \apj, 679, 101

\bibitem[Shi et al.(2006)]{shi} Shi, Y., et al. 2006, \apj, 653, 127

\bibitem[Siebenmorgen et al.(2005)]{siebenmorgen05} Siebenmorgen, R., Haas, M., Kr\"{u}gel, E., \& Schulz, B. 2005, \aap, 436, 5

\bibitem[Sirocky et al.(2008)]{sirocky} Sirocky, M. M, Levenson, N. A., Elitzur, M., Spoon, H. W. W., \& Armus, L. 2008, \apj, 678, 729

\bibitem[Soifer et al.(2000)]{soi00} Soifer, B. T., et al. 2000, \aj, 119, 509 

\bibitem[Spoon et al.(2005)]{spoon05} Spoon, H. W. W., Keane, J. V., Cami, J., Lahuis, F., Tielens, A. G. G. M., Armus, L., \& Charmandaris, V. 2005, in IAU Symp. 231, Astrochemistry: Recent Successes and Current Challenges, ed. D. C. Lis, G. A. Blake, \& E. Herbst (Cambridge: Cambridge Univ. Press), 281

\bibitem[Sturm et al.(2005)]{sturm05} Sturm, E., et al. 2005, \apj, 629, 21

\bibitem[Sturm et al.(2006)]{sturm06} Sturm, E., Hasinger, G., Lehmann, I., Mainieri, V., Genzel, R., Lehnert, M. D., Lutz, D., \& Tacconi, L. J. 2006, \apj, 642, 81

\bibitem[Teplitz et al.(2006)]{teplitz06} Teplitz, H. I., et al. 2006, \apj, 638, 1

\bibitem[Tristram et al.(2007)]{tristram07} Tristram, K. R. W., et al. 2007, \aap, 474, 837

\bibitem[Van Bemmel \& Dullemond (2003)]{van03} van Bemmel, I. M., \& Dullemond, C. P.  2003, \aap, 404, 1

\bibitem[Werner et al.(2004)]{Werner04} Werner, M. W., et al. 2004, \apjs, 154, 1


\end{thebibliography}
